\documentclass[iop]{emulateapj}
\usepackage{apjfonts}

\usepackage{ifpdf}
\usepackage{amssymb,amsmath}
\usepackage{multirow}
\usepackage{natbib}
\usepackage{color}

\def\KOI{KOI~1224}

\slugcomment{Draft \today}

\shorttitle{A Bloated Hot WD Companion Found With {\em Kepler}}
\shortauthors{Breton et al.}

\begin{document}

\title{KOI~1224, A Fourth Bloated Hot White Dwarf Companion Found With {\em Kepler}}

\author{R. P. Breton\altaffilmark{1}, S. A. Rappaport\altaffilmark{2}, M. H. van Kerkwijk\altaffilmark{1}, J. A. Carter\altaffilmark{3,4}}

\altaffiltext{1}{Department of Astronomy and Astrophysics, University of Toronto, Toronto, ON M5S 3H4, Canada; breton@astro.utoronto.ca}
\altaffiltext{2}{M.I.T. Kavli Institute for Astrophysics and Space Research, 70 Vassar St., Cambridge, MA 02139, USA}
\altaffiltext{3}{Harvard-Smithsonian CfA, 60 Garden St., Cambridge, MA 02138, USA}
\altaffiltext{4}{Hubble Fellow}

\begin{abstract}
We present an analysis and interpretation of the {\em Kepler} binary system \KOI. This is the fourth binary found with {\em Kepler} that consists of a thermally bloated, hot white dwarf in a close orbit with a more or less normal star of spectral class A or F. As we show, \KOI\ contains a white dwarf with $T_{\rm eff} = 14,400 \pm 1100$ K, mass $=\,0.20 \pm 0.02 \,M_\odot$, and radius $= \,0.103 \pm 0.004 \,R_\odot$, and an F-star companion of mass $1.59 \pm 0.07 \,M_\odot$ that is somewhat beyond its terminal-age main sequence. The orbital period is quite short at 2.69802 days. The ingredients that are used in the analysis are the {\em Kepler} binary light curve, including the detection of the Doppler boosting effect; the NUV and FUV fluxes from the {\em Galex} images of this object; an estimate of the spectral type of the F-star companion; and evolutionary models of the companion designed to match its effective temperature and mean density. The light curve is modelled with a new code named {\tt Icarus} which we describe in detail. Its features include the full treatment of orbital phase-resolved spectroscopy, Doppler boosting, irradiation effects and transits/eclipses, which are particularly suited to irradiated eclipsing binaries. We interpret the \KOI\ system in terms of its likely evolutionary history. We infer that this type of system, containing a bloated hot white dwarf, is the direct descendant of an Algol-type binary. In spite of this basic understanding of the origin of \KOI, we discuss a number of problems associated with producing this type of system with this short of an short orbital period.
\end{abstract}


\section{Introduction}
Close binary systems containing a single compact star (i.e., white dwarf, neutron star, or black hole) have the potential to enhance our knowledge of stellar evolution, perhaps even more so than for double degenerate systems whose evolution is yet more complicated. In almost all cases, the compact star originated from the remnant core of the original primary star in the binary system. The primary's envelope was lost via some combination of Roche-lobe overflow mass transfer, stellar wind, common envelope, or explosive process (e.g., a supernova).

In cases where the original secondary star is now transferring mass back to the compact star, the systems may be highly detectable via the release of gravitational potential energy in the form of optical, UV, or X-ray radiation. However, systems in which the compact star has not yet started to accrete may be much more difficult to discover. This is especially true in the case of white dwarfs in orbit with intrinsically much brighter stars that outshine them \citep[e.g., Regulus;][]{giedie08a,rappod09a}. Even in the case where the observer is in a fortuitous orientation to possibly see eclipses, such events may be of very small amplitude, e.g., at the $10^{-3}-10^{-4}$ level if the white dwarf has cooled to near the temperature of the parent star. Due to the extraordinary photometric precision of the {\em Kepler} mission, however, eclipses and transits of this depth are readily detectable.

Since the {\em Kepler} mission has been launched, there have in fact been a number of such binaries containing a white dwarf that have been discovered \citep{rowbor10a,vanrap10a,carrap11a}. In those systems the white dwarf is still sufficiently hot (i.e., $\sim 12000 - 18000$\,K) and thermally bloated (i.e., up to $\sim$ 10 times their degenerate radii) that the eclipses and transits are in the range of up to a couple of percent of the system light. We report on the study of a fourth such system with a number of similarities to the others, \KOI\ (KIC~6606653; \citealt{borkoc11a,prsbat11a}); however, it has the shortest among the orbital periods at 2.69802 days\footnote{We note that a new binary system, 1SWASP J024743.37$-$251549.2, recently discovered with the {\em WASP} survey \citep{maxand11a} contains a bloated hot white dwarf in an even shorter period binary with $P_{\rm orb} = 0.668$ days.}, and the primary star in this system has the lowest effective temperature and is the most evolved away from the main sequence of any of the previously detected hot white dwarf systems.

We report here an analysis of the {\em Kepler} Q1 and Q2 public data for \KOI , as well as supplementary observations from the {\em GALEX} satellite. We make use of stellar evolution models to better estimate the mass and evolutionary state of the current primary F star in the system.

We analyze the {\em Kepler} light curve using {\tt Icarus}, a newly developed state-of-the-art binary light curve synthesis code which is aimed at the study of detached and semi-detached binaries. This code has the advantage of handling spectroscopic data as well as Doppler boosting, irradiation effects and transits/eclipses, hence making it particularly attractive for irradiated binaries. An eclipsing binary such as \KOI\ represents an ideal testbed to benchmark the synthesis code because its results can be easily compared with those of a semi-analytic analysis. The synthesis code is described in some detail in this work.

The organization of the paper is as follows. In \S\ref{s:synthesis} we introduce the new binary light curve synthesis code. Some detailed specific aspects of the code are given in Appendix~\ref{s:icarus}. After presenting \KOI\ and the available data in \S\ref{s:data}, we proceed with the light curve modelling in \S\ref{s:modelling}, first semi-analytically and then with our synthesis code. We discuss the probable evolution of \KOI , its implications for understanding mass transfer, and its relation to Algol systems in \S\ref{s:discussion}. We summarize our findings and draw some general conclusions in \S\ref{s:conclusion}.

\section{Binary Light Curve Synthesis Code}\label{s:synthesis}
We have implemented our own light curve synthesis code with the purpose of modelling the light curves and spectra of detached and semi-detached binaries. Our code, called {\tt Icarus}\footnote{{\tt Icarus} is freely available at \url{http://www.astro.utoronto.ca/$\sim$breton/icarus}.}, is most similar to the {\tt ELC} code by \citet{orohau00a}. Essentially, the code constructs a finite-element surface grid for a star having some pre-determined physical and binary properties by solving the gravitational equipotential equation. Each surface element is characterized by a set of physical properties (i.e., temperature and an effective gravity), and the observed flux is obtained by integrating the specific intensity emerging from the surface visible by an observer located in a given direction. In the event that both stars contribute significantly to the total observed light, we model the second star of the system by inverting the mass ratio and adding a half orbital shift to it.

The core implementation is made in Python and relies heavily on the Scipy and Numpy libraries\footnote{Freely available at \url{http://scipy.org/}}. In order to bolster the execution speed, critical components are written in C and included directly in the Python code with the Scipy Weave module\footnote{See \url{http://www.scipy.org/Weave}}. The modular, object-oriented approach that we have adopted should facilitate adding features to the code. Hence, the code currently includes various effects such as eclipses/transits, irradiation from a companion, ellipsoidal light variations, and Doppler boosting. As opposed to other synthesis codes, such as the Wilson-Devinney code \citep{wildev71a}, or its modern incarnation in {\tt PHOEBE} \citep{prszwi05a}, which bundle together the binary synthesis and the parameter optimization in a single piece of software, ours resembles more a suite of routines that can be glued together in order to perform binary modelling.

The main module of our code is the physics engine, which is responsible for synthesizing the star. A separate module deals with the atmosphere model and provides utility functions that return the intensity given a set of properties -- typically the temperature, surface gravity, emission angle and, sometimes, the velocity. This layer works independently of the binary light curve physics and serves only as a frontend to retrieve the specific intensities from a synthetic atmosphere model lookup table or an analytic model (e.g., blackbody). These atmosphere models can be integrated over a passband or used with their full spectral content. In the same spirit, the physics engine does not tamper with the returned values and leaves it to the user layer to alter them if necessary (e.g., to resample the spectrum or apply reddening). The user layer exists to connect the physics engine, the atmosphere models module, and the experimental data. It typically contains (1) a data reader function; (2) a utility function that takes user input parameters, recasts them into the physics engine input parameter format, and post-processes the model data to match the observed ones; and (3) a function that returns the fit residuals or goodness-of-fit. Finally, on top of the user layer, a fitting layer can be added. The function that returns the goodness-of-fit can be fed within the framework of a minimization procedure, a brute-force search, a Bayesian optimization, or any other algorithm suited to the needs of the problem.

The main innovation of our light curve synthesis code is that it can handle phase-resolved spectroscopic light curves and hence it allows one to perform modelling without the need for fitting radial velocities separately using a template and having to introduce a correction factor for the displacement of the light-center with respect to the barycenter. This capability will be demonstrated in a forthcoming paper. Thorough details about our code are provided in Appendix~\ref{s:icarus}.

\section{Data and Data Reduction}\label{s:data}
We have obtained the \KOI\ data from the {\em Kepler} public archive\footnote{\url{http://archive.stsci.edu/kepler/publiclightcurves.html}}. The original raw light curve, presented in Figure~\ref{f:lightcurve_raw}, includes the first and second quarters (Q1 and Q2) data, which were collected using a 30-min integration time. We removed the offset between the two quarters and corrected for three jumps in the flux. Next we cleaned the raw light curve from outliers using the following method. We removed the systematic trend using a 7th-degree polynomial and folded the light curve at the orbital period. We calculated the running median and running standard deviation statistics using a 20-point window in order to identify outliers deviating by more than 1.5 standard deviations of the data scatter around the median, which we flagged and discarded for the rest of the analysis. This threshold was chosen in order to remove the obvious visible outliers as some data points were clearly beyond the regular data scatter. In total, the outliers removal discarded 699 out of the 5723 data points.

\subsection{Ultraviolet {\em GALEX} Data and Other Photometry}\label{s:galex}
In order to constrain our modelling, we started with information from the {\em Kepler} Input Catalog \citep{brow+11}, which lists optical $griz$ and infrared $JHK$ 2MASS photometry as well as estimates of the temperature $T_{\rm eff}=6352{\rm\,K}$ and reddening $E_{B-V}=0.120$ (see Table~\ref{t:photometry}). The reddening is inferred using a simple exponential dust model, which, for this source, is consistent with the total dust column of $E_{B-V}=0.114$ expected along this line of sight from the COBE/DIRBE and IRAS based dust maps of \citet{schlfd98}. Comparing the $griz$ and $JHK$ colours (dereddened using the coefficients of \citealt{schlfd98}) to those inferred from spectrophotometry of MK standards \citep{cove+07}, we infer a spectral type F6V, with an uncertainty of 1 subclass, corresponding to a temperature of $6500\pm200{\rm\,K}$ \citep{cox00}. This is consistent with that of the {\em Kepler} Input Catalog, within the expected uncertainty of $\sim\!200\,$K \citep{brow+11}. From a spectrum taken at the 1.6-m Observatoire du Mont-Megantic (L. Nelson, private communication) we infer a spectral type F5-6 III-IV (more likely luminosity class IV), which implies a temperature similar to that quoted above \citep{gragra01a}.  Below, we will use $T_{\rm eff}=6350\pm200{\rm\,K}$.

\begin{table}[!h]
\begin{center}
\caption{Archival Photometry of \KOI}
\begin{tabular}{lcccc}
\hline
\hline
Filter & Magnitude\tablenotemark{a} & Color & Reddening\tablenotemark{b} & Intrinsic \\
\hline
{\em GALEX} ${\rm FUV}$\tablenotemark{c} \dotfill & 18.98(2) & \multirow{2}{*}{1.41} & \multirow{2}{*}{-0.09} & \multirow{2}{*}{1.50} \\
{\em GALEX} ${\rm NUV}$\tablenotemark{c} \dotfill & 17.57(1) & \multirow{2}{*}{3.55} & \multirow{2}{*}{0.61} & \multirow{2}{*}{2.94} \\
KIC $g$\tablenotemark{d} \dotfill & 14.015(20) & \multirow{2}{*}{0.364} & \multirow{2}{*}{0.125} & \multirow{2}{*}{0.239} \\
KIC $r$\tablenotemark{d} \dotfill & 13.651(20) & \multirow{2}{*}{0.062} & \multirow{2}{*}{0.080} & \multirow{2}{*}{-0.018} \\
KIC $i$\tablenotemark{d} \dotfill & 13.589(20) & \multirow{2}{*}{0.033} & \multirow{2}{*}{0.073} & \multirow{2}{*}{-0.040} \\
KIC $z$\tablenotemark{d} \dotfill & 13.556(20) & \multirow{2}{*}{0.79} & \multirow{2}{*}{0.15} & \multirow{2}{*}{0.64} \\
2MASS $J$\tablenotemark{e} \dotfill & 12.77(2) & \multirow{2}{*}{0.21} & \multirow{2}{*}{0.01} & \multirow{2}{*}{0.20} \\
2MASS $H$\tablenotemark{e} \dotfill & 12.56(2) & \multirow{2}{*}{0.04} & \multirow{2}{*}{0.01} & \multirow{2}{*}{0.03} \\
2MASS $K$\tablenotemark{e} \dotfill & 12.52(3) & & & \\
\hline
\end{tabular}\label{t:photometry}
\tablenotetext{0}{The numbers in parentheses are the formal 1-$\sigma$ confidence uncertainties and are applied to the last significant figures given.}
\tablenotetext{1}{Magnitudes are in the AB system for {\em GALEX} and KIC, and Vega for 2MASS.}
\tablenotetext{2}{\citet{rey+07} for ${\rm FUV}$, ${\rm NUV}$, \citet{schlfd98} otherwise}
\tablenotetext{3}{{\em GALEX} archive (\url{http://galex.stsci.edu/})}
\tablenotetext{4}{\citet{brow+11}}
\tablenotetext{5}{\citet{skru+06}}
\end{center}
\end{table}

We checked other archives for additional information, and found that \KOI\ was detected by {\em GALEX} (see Table~\ref{t:photometry}), with $([{\rm FUV}]-[{\rm NUV}])=1.41\pm0.02$ and $([{\rm NUV}]-g)=3.55\pm0.02$. The estimated intrinsic $([{\rm NUV}]-g)_0=2.9$ color is bluer than expected for the primary ($\sim\!4.0$; \citealt{bian+07,venn+11}), but we cannot exclude that it could significantly contribute to the NUV flux given the uncertainties in the temperature, the reddening and the extinction curve. However, no contribution from the primary is expected to the FUV flux, and indeed, with $([{\rm FUV}]-[{\rm NUV}])_0=1.50\pm0.03$,\footnote{For a standard extinction curve, $A({\rm FUV,NUV})/E_{B-V}=(8.16,8.90)$ \citep{rey+07}. Given the small difference, the intrinsic color is secure.} the source is bluer in the UV than any of the brighter stellar sources near it. Thus, the FUV emission is almost certainly from the hot white dwarf secondary.

In order to estimate a temperature for the white dwarf, we can use the fact that when the white dwarf is eclipsed, the flux in the {\em Kepler} bandpass decreases by 1.4\%. Here, the 1.4\% drop includes a 10\% correction for a ``third light'' (e.g., a fainter blended star within the {\em Kepler} postage stamp), and hence it represents the white dwarf's contribution to the system's luminosity. From color terms of \citet{brow+11}, the $r$-band flux should decrease about 10\% less. We thus estimate $r\simeq18.4\pm0.2$ and infer $([{\rm FUV}]-r)\simeq0.6\pm0.2$ and $([{\rm FUV}]-r_0)\simeq0.0\pm0.3$. We can compare this with predicted {\em GALEX} and optical colors for white dwarfs from \cite{venn+11}. For their lowest mass, $0.4\,M_\odot$, they list $([{\rm FUV}]-r)=3.29,1.90,0.86,0.27,-0.03,-0.26,-0.46$ for $T_{\rm eff}=10000$ to 16000\,K in steps of 1000\,K, with the color depending somewhat on gravity below 13000\,K (here, we converted the V-band magnitudes of \citet{venn+11} to $r$-band using $V-r=0.19$, valid for temperatures in this range; \citealt{fuku+96}). Thus, we infer a temperature of $\sim\!14000\pm1000\,$K.

For the above temperature, the models also predict $([{\rm FUV}]-[{\rm NUV}])_0=0.0\pm0.1$, which is much bluer than the observed color. Apparently, therefore, either our estimate is inaccurate and the true temperature is $10500\pm500{\rm\,K}$, or the primary contributes $\sim\!75\%$ of the NUV flux, i.e., is about a magnitude brighter than expected (perhaps because of stellar activity associated with rapid rotation). From the ratio of the transit to eclipse depths (see \S\ref{s:analytic}), it seems clear the latter is the correct interpretation, and the white dwarf effective temperature is $\sim 14000 \pm 1000\,$K. This is further confirmed by our formal fits to the light curve (see \S\ref{s:numeric}).

\section{Modelling the Light Curve of \KOI}\label{s:modelling}
\subsection{Semi-Analytic Modelling}\label{s:analytic}
\begin{figure*}[!h]
\centerline{\hfill
\ifpdf
    \includegraphics[width=0.7\hsize]{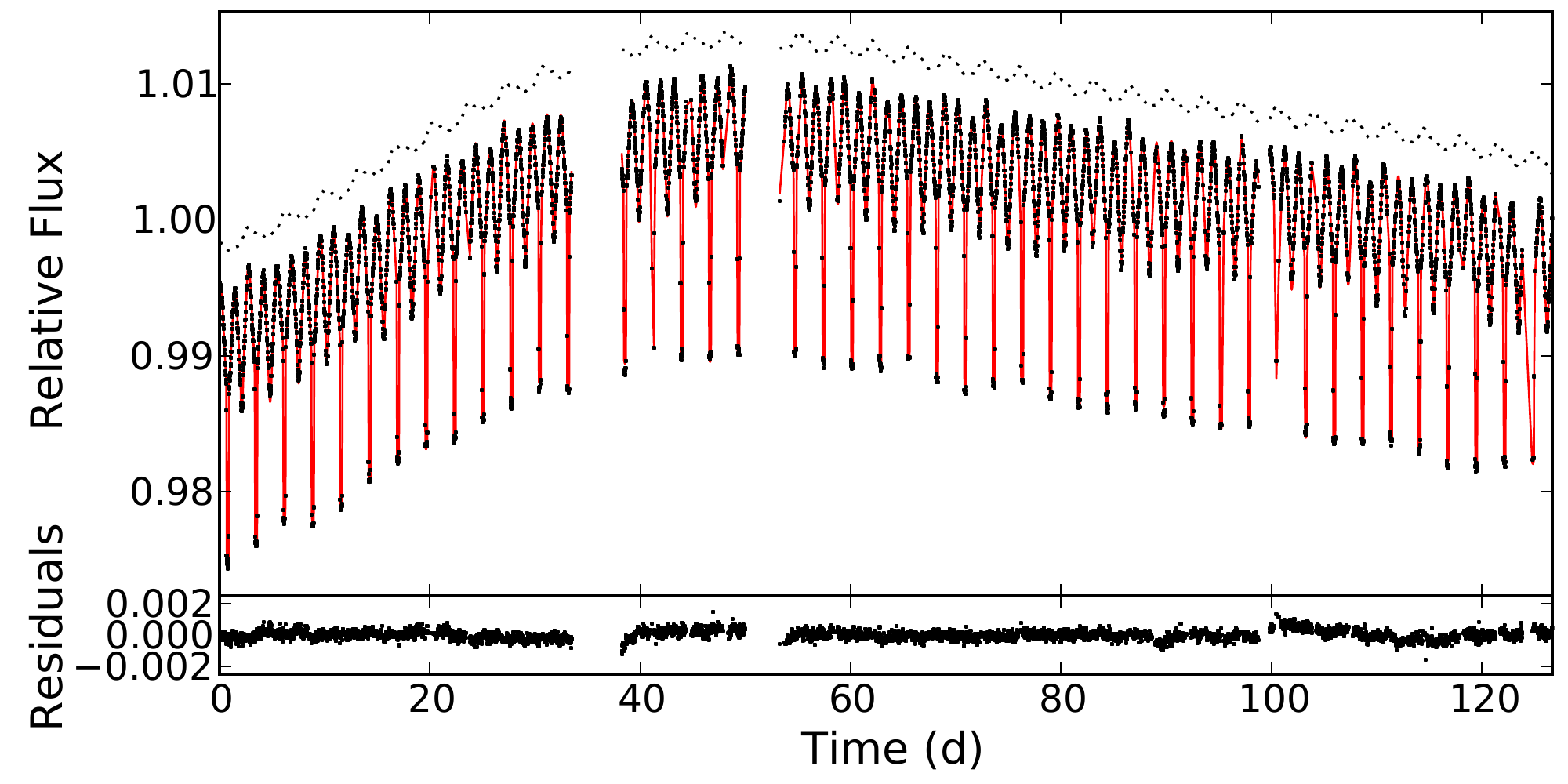}%
\else
    \includegraphics[width=0.7\hsize]{lightcurve_raw.eps}%
\fi
\hfill}
\caption{Raw light curve of \KOI\ from {\em Kepler} (black dots) with a best-fit solution (solid line). The dashed line shows the 7th-degree polynomial and the three pulsation harmonics used to remove the systematic trend (shifted upward for visibility reasons). The lower panel displays the fit residuals.\label{f:lightcurve_raw}}
\end{figure*}

We first performed a semi-analytical analysis of the \KOI\ light curve in order to obtain rough estimates of the system parameters. We fitted the raw data (see Figure~\ref{f:lightcurve_raw}) for three out-of-eclipse orbital harmonics as well as a 7th-degree polynomial and three ``pulsations'' (at 3.49, 1.75 and 1.17\,d). The polynomial function aims to remove the long-term instrumental variations, while the pulsations were detected as significant signals in a power spectrum. The 3.49-day pulsation is strongest and might reflect the rotation period of the primary, as seemed to be the case for KOI~74, where a possible pulsation period of $\sim\!0.6{\rm\,d}$ matched the rotation period inferred from its rotational velocity.  If so, it would imply the star is rotating sub-synchronously, perhaps because the star is expanding during the course of its evolution. We then removed the systematic trend from the data and folded them at the orbital period.

\begin{figure*}[!h]
\centerline{\hfill
\ifpdf
    \includegraphics[width=0.7\hsize]{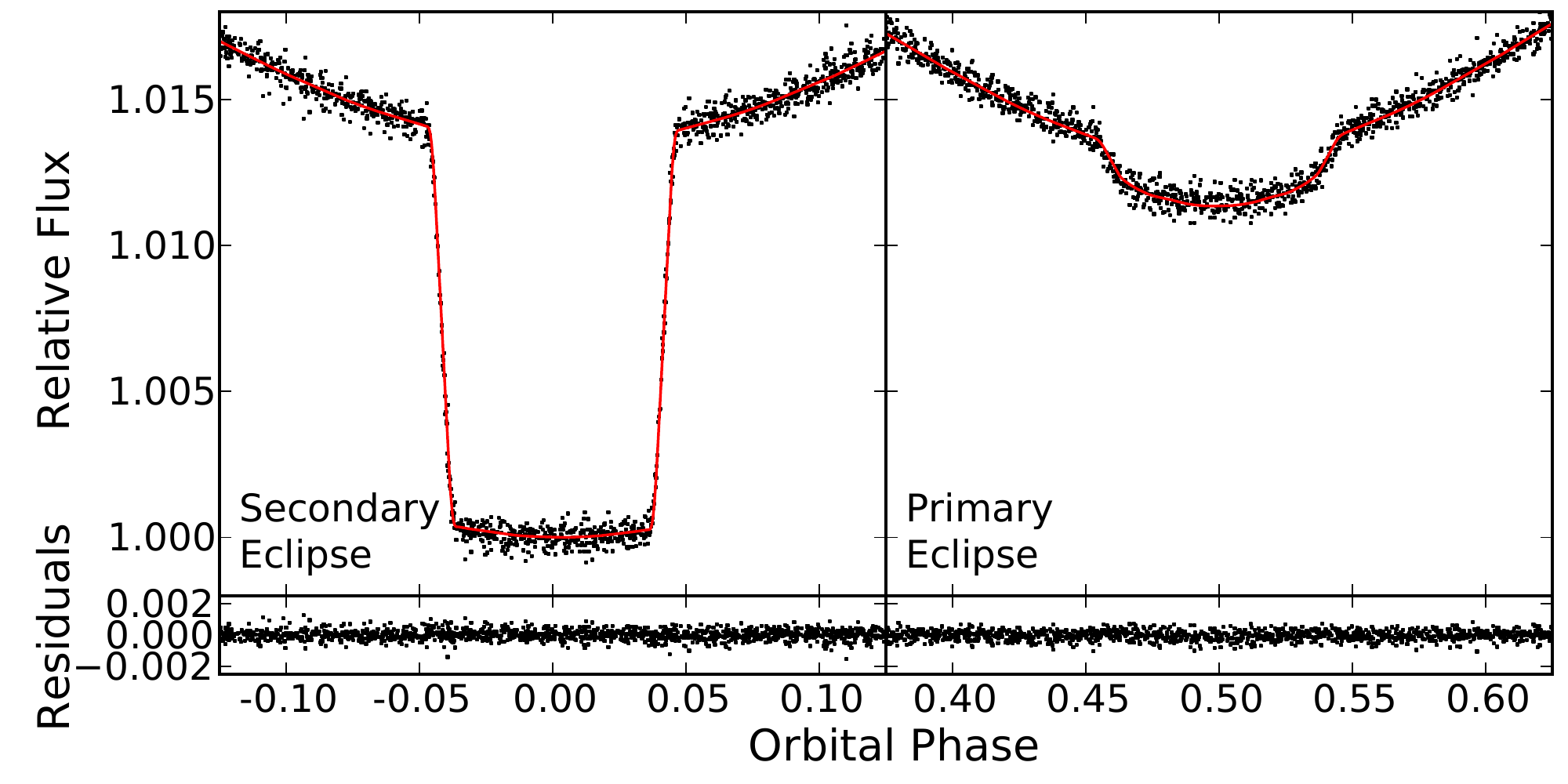}%
\else
    \includegraphics[width=0.7\hsize]{lightcurve_eclipses.eps}%
\fi
\hfill}
\caption{Light curve of \KOI \, (black dots) along with a best-fit solution (solid line) folded at the orbital period of the system. The systematic trend and the third light contribution have been removed from the data, and the flux range is adjusted to display the primary and the secondary eclipses in the right and left panels, respectively. The lower panels show the fit residuals.\label{f:lightcurve_eclipses}}
\end{figure*}

\begin{figure*}[!h]
\centerline{\hfill
\ifpdf
    \includegraphics[width=0.7\hsize]{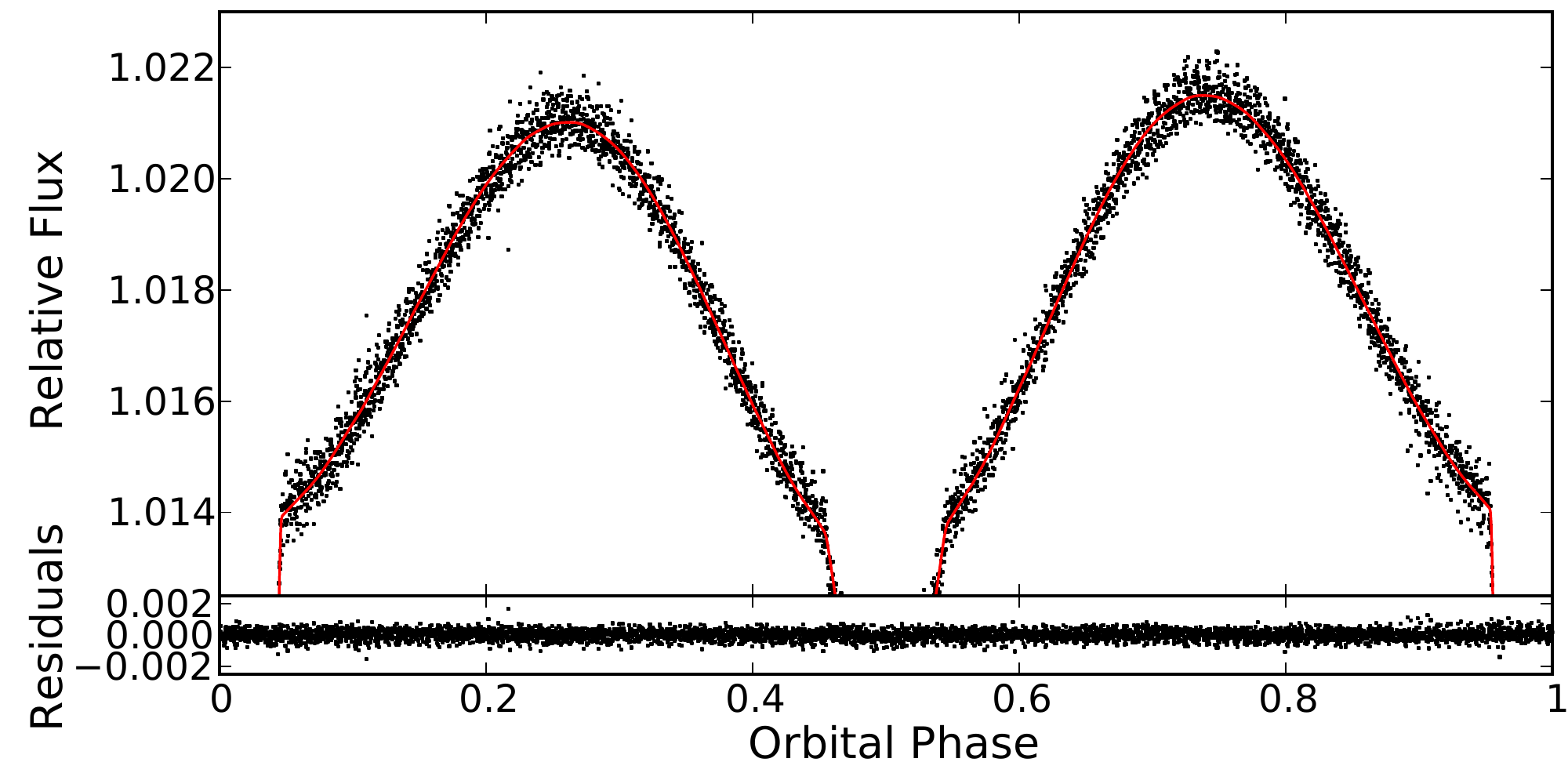}%
\else
    \includegraphics[width=0.7\hsize]{lightcurve_folded.eps}%
\fi
\hfill}
\caption{Light curve of \KOI \,(black dots) along with a best-fit solution (solid line) folded at the orbital period of the system. The systematic trend and the third light contribution have been removed from the data, and the flux range is adjusted to trim the eclipses in order to enhance the visibility of the ellipsoidal and Doppler boosting components. The lower panel displays the fit residuals.\label{f:lightcurve_folded}}
\end{figure*}

We measured various fiducial points on the white dwarf eclipse profile using a detailed plot of the light curve (Figure~\ref{f:lightcurve_eclipses}) in order to estimate the radius of the primary star in units of the semi-major axis, i.e., $R_1/a$. We found the phase of the first contact, the start of the full eclipse, the end of the full eclipse, and the last contact to be $\phi_1 \sim -0.0471$, $\phi_2 \sim -0.0369$, $\phi_3 \sim 0.0365$ and $\phi_4 \sim 0.0466$, respectively. From these values, measured in orbital cycles from the primary's inferior conjunction, we can infer $R_1/a = \pi (\phi_3 - \phi_1) = 0.263$ and $R_1/a = \pi (\phi_4 - \phi_2) = 0.262$ (implicitly assuming $i = 90^\circ$, since we have no other way to infer it from simple geometry).

The first three out-of-eclipse orbital harmonics capture information about the system properties that can be derived from ellipsoidal light variations, irradiation effects, and Doppler boosting (see the folded out-of-eclipse light curve in Figure~\ref{f:lightcurve_folded}). From our fit we found the fractional amplitudes $\mathcal{A}_1=(2.542 \pm 0.058)\times10^{-4}$, $\mathcal{A}_2=(37.675 \pm 0.071)\times10^{-4}$ and $\mathcal{A}_3=(3.085 \pm 0.074)\times10^{-4}$ along with the corresponding phases $\phi_1=(-6.25 \pm 0.48)\times10^{-2}$, $\phi_2=(2.6 \pm 1.1)\times10^{-4}$ and $\phi_3=(1.5 \pm 1.0)\times10^{-4}$. The phases are in orbital cycles measured from the ascending node of the white dwarf for $\phi_{1,2}$ and from the white dwarf eclipse for $\phi_3$. Because the ingress and egress time scales are comparable to that of the {\em Kepler} integration time, it would be futile to make any inference about $R_2/a$ without a proper numerical treatment, as in \S\ref{s:numeric}. 

In the simple, semi-analytic analysis that follows, we correct amplitudes $\mathcal{A}_1$ and $\mathcal{A}_2$ {\em up} by a factor of $\sim$10\% to account for the third light in the system, and $\mathcal{A}_1$ {\em up} by an additional $\sim$6\% to account for the much smaller expected Doppler boosting effect of the white dwarf which {\em decreases} the total amplitude of $\mathcal{A}_1$ \citep[see, e.g.,][equation 11]{carrap11a}. We label these ``corrected'' amplitudes $\mathcal{A}'_1$ and $\mathcal{A}'_2$.

The first out-of-eclipse orbital harmonic arises from Doppler boosting (see \S\ref{s:doppler}, and \citealt{vanrap10a,loegau03a} for more details) and has an amplitude:
\begin{equation}
    \mathcal{A}'_1 \cos \phi_1 = f_{\rm DB} \frac{v}{c} \,.
\end{equation}
From the color information of \KOI\ presented in \S\ref{s:galex}, we estimate $f_{\rm DB} = 3.55$ at 6350\,K and $\log g = 4$. Using the above expression we find that the harmonic amplitude $\mathcal{A}'_1$ implies a projected velocity amplitude $K_1 \simeq 23 \pm 1$\,km/s, where the cited uncertainty includes 3\% each from the amplitude and $f_{\rm DB}$. From $K_1$ we find the mass function:
\begin{equation}
\frac{M_2^3 \sin^3 i}{\left(M_1+M_2\right)^2} \simeq 0.00344 \pm 0.00040
\end{equation}
The second orbital harmonic is due to the ellipsoidal light variations, and has an expected amplitude:\begin{equation}
    \mathcal{A}'_2 \cos\phi_2 = f_{\rm EV} \frac{M_2}{M_1}\left(\frac{R_1}{a}\right)^3\sin^2 i, \label{eq:ellipsoidal}
\end{equation}
where $f_{\rm EV}$ is a pre-factor of order unity that depends on the limb and gravity darkening:
\begin{equation}
    f_{\rm EV} = \frac{45 + 3u}{20(3 - u)} (\tau + 1)
\end{equation}
\citep{mor85a}\footnote{Here we dropped the factor $k$ from the original equation because it is negligible}. Using the corrected amplitude, $\mathcal{A}'_2$ and the geometrically measured value of $R_1/a$, and considering that $\sin i \sim 1$, we find a mass ratio $q = M_2/M_1 = (0.23 \pm 0.02)/f_{\rm EV}$. We also find that $f_{\rm EV} \simeq 1.59$ if we take the linear limb darkening $u \sim 0.56$ for Kepler's bandpass at $6350$\,K from \citet{sin10a}, and $\tau \sim 0.66$ by integrating Equation~10 from \citet{mor85a} over the Kepler bandpass. Hence, we estimate that $q = 0.144 \pm 0.020$. One also sees that the ellipsoidal term is phased as expected with the ascending node.

Combining the mass function found from the $\mathcal{A}'_1$ amplitude and the mass ratio from the $\mathcal{A}'_2$ term, we find
\begin{equation}
    M_1 \simeq 1.5\,M_\odot~~~~~M_2 \simeq 0.22 \, M_\odot \,.
\end{equation}

Second order terms from ellipsoidal light variations arise because the nose of the star (i.e., near the L$_1$ point) is darker than its back side. This results in the third harmonic, with a relative amplitude $\mathcal{A}_3/\mathcal{A}_2 = 0.082 \pm 0.002$. This values differs slightly with that of 0.06 calculated with the equations of \citet{mor85a}, assuming the same coefficient as above.

Lastly, there should also be a contribution to the fundamental from the ellipsoidal light variation effect, with an amplitude of $3\mathcal{A}'_3/5 = (2.04 \pm 0.05) \times 10^{-4}$, with a maximum at the white dwarf's eclipse. Instead, the fundamental modulation is a bit shifted towards the white dwarf transit, with amplitude $\mathcal{A}'_1 \sin(\phi_1) = (1.07 \pm 0.08) \times 10^{-4}$. This indicates that the shift of the fundamental modulation is mainly due to irradiation, rather than second-order ellipsoidal light variations. Assuming the second-order contribution due to the ellipsoidal light variation is correct, we infer an irradiation term of $f_{\rm irr} \sim (3.11 \pm 0.09) \times 10^{-4}$. According to harmonic decomposition of the reflection effects in close binaries by \citet{kop59a}, we can write the fractional amplitude of the first harmonic term due to reflection as:
\begin{equation}
    f_{\rm irr} \simeq \alpha_1 \frac{L_2}{L} \left( \frac{R_1}{a} \right)^2 \left[ \frac{1}{3} + \frac{1}{4}\left(\frac{R_1}{a}\right) \right] \,,
\end{equation}
where $\alpha$ is the primary's albedo, $L_2/L$ is the ratio of the secondary's luminosity to the total luminosity. In this estimate we have ignored the leading-order $(R_2/a)^2$ term (due to the irradiation of the secondary by the primary) because it is nearly two orders of magnitude smaller than the $L_2/L(R_1/a)^2$ term -- as we find in \S\ref{s:numeric} (see Table~\ref{t:parameters}). From this we infer $L_2/L \sim 0.011\alpha$, which is quite consistent with the $\sim$$1.4\%$ drop in the light curve during the white dwarf eclipse.  

In the next section we discuss our numerical modelling from which we obtain more accurate results for the system masses. The main reason for this improved accuracy is that we utilze measured properties of the primary star to directly determine its mass. In that case, the use of $\mathcal{A}'_2$ in determining the mass ratio is less critical, and, more importantly, $M_2$ then depends principally on the cube root of the mass function, and so is only linearly dependent on $\mathcal{A}'_1$ (rather than on its cube).

\subsection{Numerical Modelling}\label{s:numeric}
The numerical modelling has been accomplished using our new binary light curve synthesis code {\tt Icarus} sketched in \S\ref{s:synthesis} and described more completely in Appendix~\ref{s:icarus}. 

In order to fully describe the light curves, our code includes the following list of free parameters (with indices 1 and 2 referring to the F-star primary and the white dwarf secondary, respectively), which are also summarized in Table~\ref{t:parameters}: the mass ratio, $q = M_2/M_1$; the filling factor\footnote{See \S\ref{s:equipotential} for definition.}, $f_{1,2} = x_{{\rm nose}\,(1,2)} / x_{L1\,(1.2)}$; the (polar) temperature ratio, $T_2 / T_1$; the primary's (polar) temperature, $T_1$; the irradiation temperature ratio, $T_{1,{\rm irr}}/T_1$ (see \S\ref{s:irradiation} for the definition of $T_{1,{\rm irr}}$); the orbital inclination, $i$; and a third light contribution, $L_3$.

\begin{table}[!h]
\begin{center}
\caption{\KOI\ Parameters}
\begin{tabular}{lr}
\hline
\hline
Parameter & Value \\
\hline
Orbital period, $P_{\rm orb}$ (d) \dotfill & 2.69802(2) \\
Epoch of primary's inferior conjunction, $T_0$ (MJD) \dotfill & 55030.038(1) \\
\hline
\multicolumn{2}{c}{Model Parameters} \\
\hline
Mass ratio, $q = M_2/M_1$ \dotfill & 0.128(10) \\
Filling factor, $f_1$ (fraction of $x_{L1}$\tablenotemark{a}) \dotfill & 0.389(6) \\
Filling factor, $f_2$ (fraction of $x_{L2}$\tablenotemark{a}) \dotfill & 0.034(2) \\
Temperature, $T_1$ (K) \dotfill & 6250(200) \\
Temperature ratio, $T_2/T_1$ \dotfill & 2.3(1) \\
Temperature ratio, $T_{\rm irr}/T_1$ \dotfill & 0.164(3) \\
Orbital inclination, $i$ (degree) \dotfill & 85(1) \\
Third-light, $L_3$ (fraction of total light) \dotfill & 0.10(5) \\
Extra noise, $\sigma_{\rm extra}$ (fraction of average flux errors) \dotfill & 3.69(4) \\
\hline
\multicolumn{2}{c}{Derived Parameters} \\
\hline
Mass primary, $M_1$ ($M_\odot$) \dotfill & 1.59(7) \\
Mass secondary, $M_2$ ($M_\odot$) \dotfill & 0.20(2) \\
Semi-major axis, $a$ ($R_\odot$) \dotfill & 9.9(2) \\
Projected velocity amplitude, $K_1$ (km\,s$^{-1}$) \dotfill & 21.0(1.7) \\
Temperature, $T_2$ (K) \dotfill & 14400(1100) \\
Relative volume-averaged radius, $\langle R_1 \rangle/a$ \dotfill & 0.270(5) \\
Relative volume-averaged radius, $\langle R_2 \rangle/a$ \dotfill & 0.0103(4) \\
Volume-averaged radius, $\langle R_1 \rangle$ \dotfill & 2.67(6) \\
Volume-averaged radius, $\langle R_2 \rangle$ \dotfill & 0.103(4) \\
Luminosity, $L_1$ ($L_\odot$) \dotfill & 9.8(1.3) \\
Luminosity, $L_2$ ($L_\odot$) \dotfill & 0.41(13) \\
Volume-average density, $\langle \rho_1 \rangle$ (g\,cm$^{-3}$) \dotfill & 0.118(6) \\
Volume-average density, $\langle \rho_2 \rangle$ (g\,cm$^{-3}$) \dotfill & 270(30) \\
Polar surface gravity, $\log g_1$ ($\log_{10}({\rm cm\,s}^{-2})$) \dotfill & 3.79(2) \\
Polar surface gravity, $\log g_1$ ($\log_{10}({\rm cm\,s}^{-2})$) \dotfill & 5.72(5) \\
\hline
\end{tabular}\label{t:parameters}
\tablenotetext{0}{The numbers in parentheses are the formal 1-$\sigma$ confidence uncertainties and are applied to the last significant figures given.}
\tablenotetext{1}{See \S\ref{s:equipotential} for definition.}
\end{center}
\end{table}

\begin{figure}[!h]
\centerline{\hfill
\ifpdf
    \includegraphics[width=0.9\hsize]{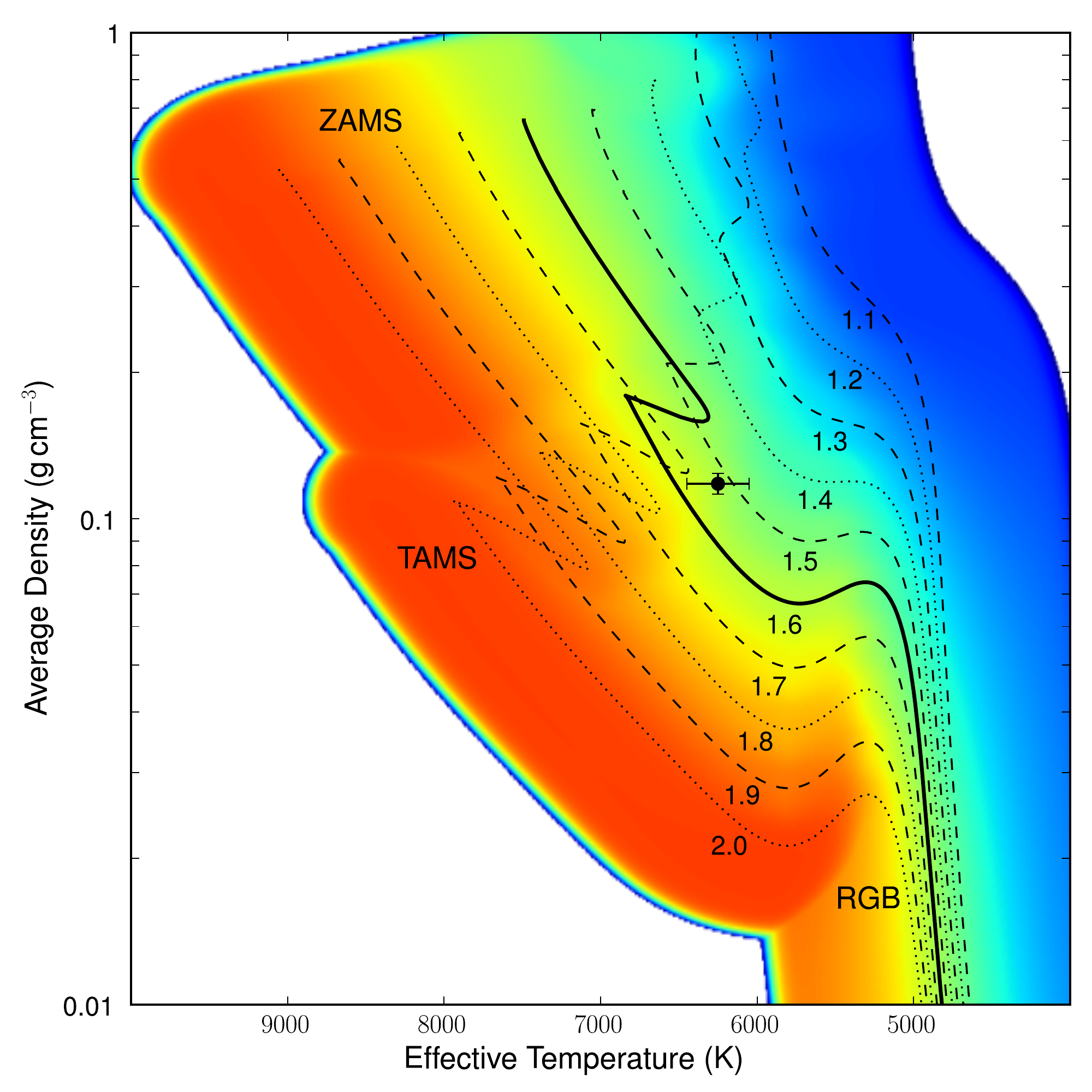}%
\else
    \includegraphics[width=0.9\hsize]{rho_teff.eps}%
\fi
\hfill}
\caption{Evolution of the primary star in the $T_{1, {\rm eff}}-\langle \rho_1 \rangle$ plane. The 10 curves are for stars starting on the ZAMS for masses between 1.1 and 2.0 $M_\odot$ in steps of 0.1 $M_\odot$. The colored region represents a nearly continuous mapping of primary mass in the $T_{1, {\rm eff}}-\langle \rho_1 \rangle$ plane.  This map was generated from the evolution tracks using an assumed 2-dimensional Gaussian probability distribution with 1-$\sigma$ uncertainties of 200 K in $T_{1, {\rm eff}}$ and 10\% in $\langle \rho_1 \rangle$. The black dot indicates the position of \KOI\ from our numerical modelling along with the 68\% confidence intervals on the temperature and density.\label{f:rho_teff}}
\end{figure}

One more parameter is required in order to fully specify the system dynamics and solve for the masses. This is required in order to determine the projected velocity amplitude of the primary, $K_1$, which, in turn, allows us to calculate the Doppler boosting amplitude. Instead of introducing this quantity as an additional free parameter to the fit, we use the primary mass, $M_1$, itself; this is possible because the mean stellar density of the primary, $\langle \rho_1 \rangle$, and its effective temperature $T_{\rm eff}$\footnote{For this purpose, we approximated the effective temperature by $T_1$, the polar temperature.} determine a nearly unique mass and age, provided that the star has not evolved far up the giant branch. The average density of the primary star, in turn, can be found using Kepler's third law, which depends only on known quantities such as the volume averaged stellar radius $\langle R_1 \rangle /a~$\footnote{$\langle R_1 \rangle /a$ is found after solving for the equipotential surface, which does not require knowing the masses in the system.}:
\begin{equation}
    \langle \rho_1 \rangle = \frac{3 \pi}{G P_{\rm orb}^2 (1+q)} \left(\frac{a}{\langle R_1 \rangle} \right)^3 \,.
\end{equation}

We have used the newly developed {\tt MESA} stellar evolution code \citep{paxbil11a} to evolve a sequence of stars between 1.1 and 2.0\,$M_\odot$ in steps of 0.1\,$M_\odot$. These tracks are shown in Figure~\ref{f:rho_teff} in the $\langle \rho_1 \rangle - T_1$ plane. Given that there is an uncertainty in both $\langle \rho_1 \rangle$ and $T_1$, and the fact that the evolution tracks cross each other near the TAMS, we have converted this discrete set of tracks to a nearly continuous mass distribution in the  $\langle \rho_1 \rangle-T_1$ plane (see Figure~\ref{f:rho_teff}). To generate this finely gridded ($400 \times 600$) mass array we generated a probability weighting function associated with each point on the evolution tracks from a 2-dimensional Gaussian distribution in $\langle \rho_1 \rangle$ and $T_1$ -- as described in the caption to Figure~\ref{f:rho_teff}. If more than one probability was assigned to a given point in the $\langle \rho_1 \rangle - T_1$ plane for any given track, we took the highest of the probabilities. Finally, we used these probabilities to compute a weighted ``mean'' mass at each location in the $\langle \rho_1 \rangle - T_1$ plane.

We fixed the orbital ephemerides at the values reported in Table~\ref{t:parameters} and, because the rotational periods of the stars are unknown, we have assumed that they are rotating co-synchronously with the orbit for the calculation of the equipotential surfaces. We also used a gravity darkening coefficient $g_{\rm dark} = 0.175$ for both stars based on empirical values from \citet{chemon11a}\footnote{This is appropriate for the primary star of $T_1 = 6350$K, and the hotter white dwarf is sufficiently small that its gravity darkening has a negligible effect.}. As we pointed out in \S\ref{s:analytic}, irradiation effects on the white dwarf due to the primary are negligible ($\sim$2\% of the primary's irradiation by the secondary) and hence can be safely neglected. From exploratory fitting, we also realized that the flux uncertainties were insufficient to account for the root-mean-square of the fit residuals (i.e., the $\chi^2$ was large even though the fits visually looked good).  Hence, we introduced an extra noise parameter, $\sigma_{\rm extra}$, which was added in quadrature to the flux uncertainties. 

The process of calculating the goodness of fit of our 9-parameter model works as follows. We evaluate the synthetic light curve using a variable sampling scheme (i.e., more densely sampled near the transit/eclipse and less densely in the smoother ellipsoidal light variations region). This allows us to boost the computing speed while still capturing the critical details. The 30-min integration time of {\em Kepler} produces a non-negligible smearing that needs to be accounted for. Hence, we resample the synthetic light curve at a higher resolution using a second-order spline in order to perform a 30-min boxcar filtering. The synthetic light curve is then interpolated at the phase of each observed data point.  Next, the systematic trend in the data is removed after fitting a 7th-order polynomial and three sinusoidal pulsations having unknown phases, amplitudes and harmonically related periods $P_{\rm puls.}$, $P_{\rm puls.}/2$ and $P_{\rm puls.}/3$, where the fundamental is $\sim$$3.49$\,d. Finally, the residuals are calculated using the detrended light curve.

\subsubsection{MCMC Algorithm}\label{s:mcmc}
We performed the parameter optimization using a Markov Chain Monte Carlo (MCMC) algorithm since it is well suited to making parameter inferences in high-dimensional problems. Parameter correlations are sometimes complicated and hence we used a Gibbs sampler in our MCMC because it allows for an easier tuning of the proposal distribution's acceptance rate \citep[see e.g.][for more information about MCMC and Gibbs sampling]{wals2004,gre05b}.

We sampled the orbital inclination using flat priors in $\cos i$, such that it eliminates projection effects. Based on the {\em Kepler} Input Catalog estimation of the third light (see \S\ref{s:galex}), which is listed as $\sim 10\%$, we used a logistic prior for this parameter:
\begin{equation}
    \left[1 + \exp \left((L_3-0.15)/0.01\right) \right]^{-1}\,.
\end{equation}
In such way, the probability of $L_3$ values in the range $0-0.1$ is close to unity before dropping to 50\% at 0.15, and then becomes negligible beyond 0.2. For the extra noise parameter, we used modified Jeffrey's priors $[\sigma_{\rm extra} + 0.1]$, which provides equal weighting per decade. This choice is motivated by the scaling nature of the extra noise parameter. The constant of 0.1 linearizes the prior when $\sigma_{\rm extra}$ becomes smaller than 10\% of the average flux measurement errors. In a Bayesian parameter estimation framework, the extra noise parameter has an effect similar to normalizing the reduced $\chi^2$ to unity in standard $\chi^2$ fitting. Since {\em Kepler} uses a single broadband filter and the data contain no color information, we also made use of the photometry constraints (see \S\ref{s:galex}) to add a Gaussian prior of $6350 \pm 200$\,K on the primary's temperature. Flat priors were assumed for the remaining parameters.

\begin{figure*}[!h]
\centerline{\hfill
\ifpdf
    \includegraphics[width=0.9\hsize]{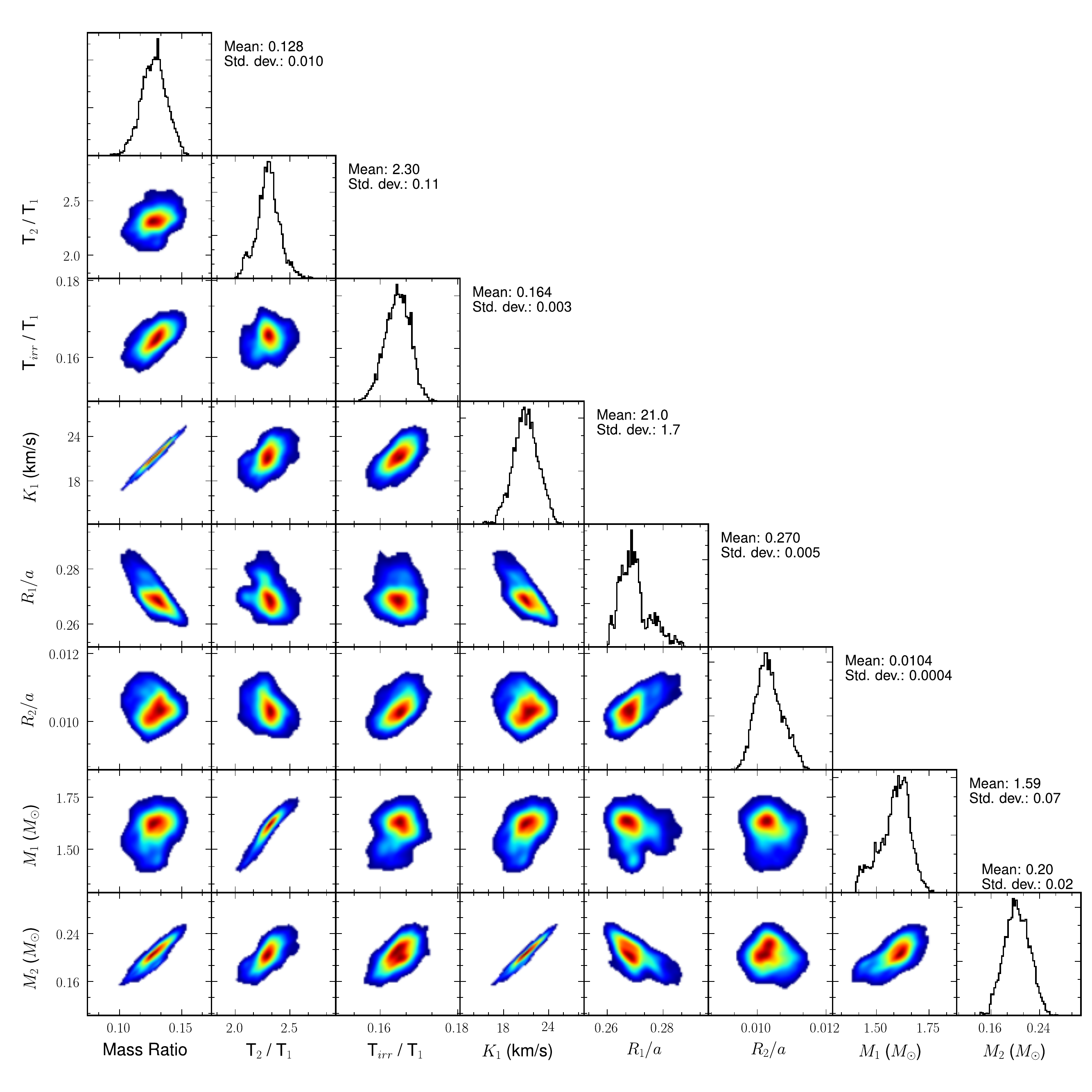}%
\else
    \includegraphics[width=0.9\hsize]{mcmc.eps}%
\fi
\hfill}
\caption{Results of the MCMC parameter estimation for the \KOI\ light curve. The data from 6 independent chains were combined after applying a burn-in of 20,000 steps and a thinning of 360 iterations. Note that $K_{1} $, $R_1/a$, $R_2/a$, $M_1$ and $M_2$ are all inferred values from other model parameters (see \S\ref{s:modelling} for more details).\label{f:mcmc}}
\end{figure*}

We ran a total of 6 independent MCMC chains, each consisting of 400,000 steps. A burn-in period of 20,000 steps was removed from each chain and we kept only every 360 iterations -- a process called thinning -- in order to reduce auto-correlation effects. We compared the summary statistics of each MCMC to the others in order to ensure that convergence has been reached. For all model parameters, the Gelman-Rubins convergence diagnostic \citep{gelm92a} yielded $\sqrt{\hat R} < 1.1$, which indicates that the chains have reached stationary distributions. The final results presented in Table~\ref{t:parameters} and Figure~\ref{f:mcmc} were obtained by merging all the chains together (after burn-in and thinning), while Figures~\ref{f:lightcurve_eclipses} and \ref{f:lightcurve_folded} display an example of the best-fit light curve.

We find constituent masses of $1.59 \pm 0.07 \, M_\odot$ and $0.20 \pm 0.02 \, M_\odot$ for the primary and white dwarf, respectively.  The radius of the primary is $2.67 \pm 0.06 \, R_\odot$, indicating that it has evolved substantially off the zero-age main sequence, consistent with its low mean density of 0.12 g cm$^{-3}$.  The white dwarf's radius, $0.103 \pm 0.004 \, R_\odot$, is $\sim$5 times its degenerate radius, and indicates that it is still actively cooling (possibly punctuated with episodic shell flashes). In turn, this indicates that its cooling age is of order 120 -- 600 Myr for masses in the range of $0.20-0.22 \, M_\odot$\citep[see, e.g.,][]{driblo99a,neldub04a}.  We discuss below what these properties may imply for the formation of \KOI.

\section{The Origin of \KOI}\label{s:discussion}
We have presented results for a system found with the {\em Kepler} mission that contains a thermally bloated hot white dwarf: \KOI\ (KIC~6606653). The system consists of a normal F6 star that has evolved to near the TAMS in a 2.698-day binary with a white dwarf of mass $0.2 \,M_\odot$ and radius $0.1 \, R_\odot$. We discuss in this section how such a binary may have evolved.

There are two plausible evolutionary paths to the formation of the current \KOI\ binary system. The transfer of mass from the white dwarf progenitor to the primordial secondary was either dynamically {\em unstable}, leading to a common envelope phase, or it was {\em stable}, even if not completely conservative. The former case leads to a dramatic shrinkage of the orbital separation while the latter results in only modest changes in the orbit. We consider each of these, in turn, and in more detail. For more discussion of the formation and evolution of these systems see \citet{farm03a}, \citet{dist11a}, \citet{vanrap10a} and \citet{carrap11a}.

\subsection{Common Envelope Scenario}
In the common-envelope scenario, the primary star develops a degenerate He core of mass $\sim$$0.2\,M_\odot$, starts to transfer mass to the secondary, and initiates a dynamically unstable process leading to a common-envelope phase. The secondary spirals into the common envelope of the primary, ejects the envelope, and this results in a very close orbit of the secondary with the He core of the primary (i.e., the current white dwarf of the system). The final orbital separation can be related to the initial (pre-CE) separation by 
\begin{equation}
    a_f \simeq a_i ~\frac{M_c M_s}{M_p M_e} \, \frac{r_L}{2} \lambda \, \alpha_{\rm CE}
\end{equation}
\citep[see, e.g.][]{dek90a,podrap03a} where $M_c$, $M_e$, and $M_p$ are the core, envelope, and total mass of the primordial primary star, respectively, $M_s$ is the mass of the secondary, $r_L$ is the Roche-lobe radius of the primary in units of the orbital separation, and the product $\lambda \alpha_{\rm CE}$ encapsulates the energy ejection efficiency and binding energy of the common envelope. The latter CE parameter is often taken to be $\sim$1, but is more likely to $\lesssim 0.1$ for an evolved star \citep[see, e.g.][]{dewtau00a,taudew01a,podrap03a}. In this latter case, the orbit must shrink by a factor of $\gtrsim$100. For a final orbital period of 2.7 days, the initial period would have to have been $\gtrsim 5$ years. Such a wide initial orbit for the progenitor binary would necessarily imply the development of a much more massive He (or likely CO) white dwarf of $\gtrsim 0.5\,M_\odot$. Thus, given the fact that the current white dwarf in \KOI\ is much lower in mass, this seems to essentially rule out a common-envelope scenario.

\subsection{Stable Mass-Transfer Scenario}
Stable, but not necessarily conservative, mass transfer from the progenitor of the white dwarf to the secondary seems the much more likely route to the production of \KOI, though, as we discuss, this formation scenario is not without its own difficulties. In this scenario, when the primary fills its Roche lobe and commences mass transfer to the secondary, that mass transfer will proceed on the thermal timescale of the envelope of the donor star \citep[see, e.g.,][]{podrap02a,linrap11a} until well after the masses of the two stars have become equal. Thermal timescale mass transfer results since the donor star is more massive than the accretor and it is not too evolved. The remainder of the mass transfer is driven by the nuclear evolution of the primary and is generally much slower than the thermal timescale transfer \citep{linrap11a}.

Depending on the relative masses of the primary and secondary at the time mass transfer commences, the outer envelope of the accreting star may swell up to the point where further accretion is inhibited and a substantial fraction of the transferred mass is ejected from the system \citep[see, e.g.][]{kipmey77a,vandeg10a,vandeg11a}. If we define the mass retention fraction by the accretor to be $\beta$, the value of $\beta$ is likely to be low at the start of the mass transfer process and higher toward the end.

As we have shown, the age of the white dwarf in \KOI\ is likely to be $\lesssim 600$ Myr because it is still quite hot \citep[see, e.g.][]{driblo99a,neldub04a}. The fact that the current primary star in \KOI\ has a mass of $\sim$1.6 $M_\odot$ and has evolved to at least the TAMS (and probably somewhat beyond) implies an age of $\gtrsim 1.75\pm 0.3$ Gyr for a single, isolated star of that mass. These two facts taken together imply that the current primary must have been already substantially evolved at the end of the mass-transfer phase when the white dwarf was unveiled. From all of this we can infer that the primordial secondary (now the current primary) was not substantially less massive than the primordial primary.

Therefore, while there likely was a brief phase of thermal timescale mass transfer, the bulk of the mass transfer should have been at a slower rate and therefore could have been mostly conservative. For purposes of simplicity and for presenting illustrative examples, we therefore parameterize the evolution of \KOI\ as having a constant mass retention fraction, $\beta$. We further take the specific angular momentum of any ejected material to be $\alpha$ in units of that of the binary system.

\begin{figure}[!h]
\centerline{\hfill
\ifpdf
    \includegraphics[width=0.9\hsize]{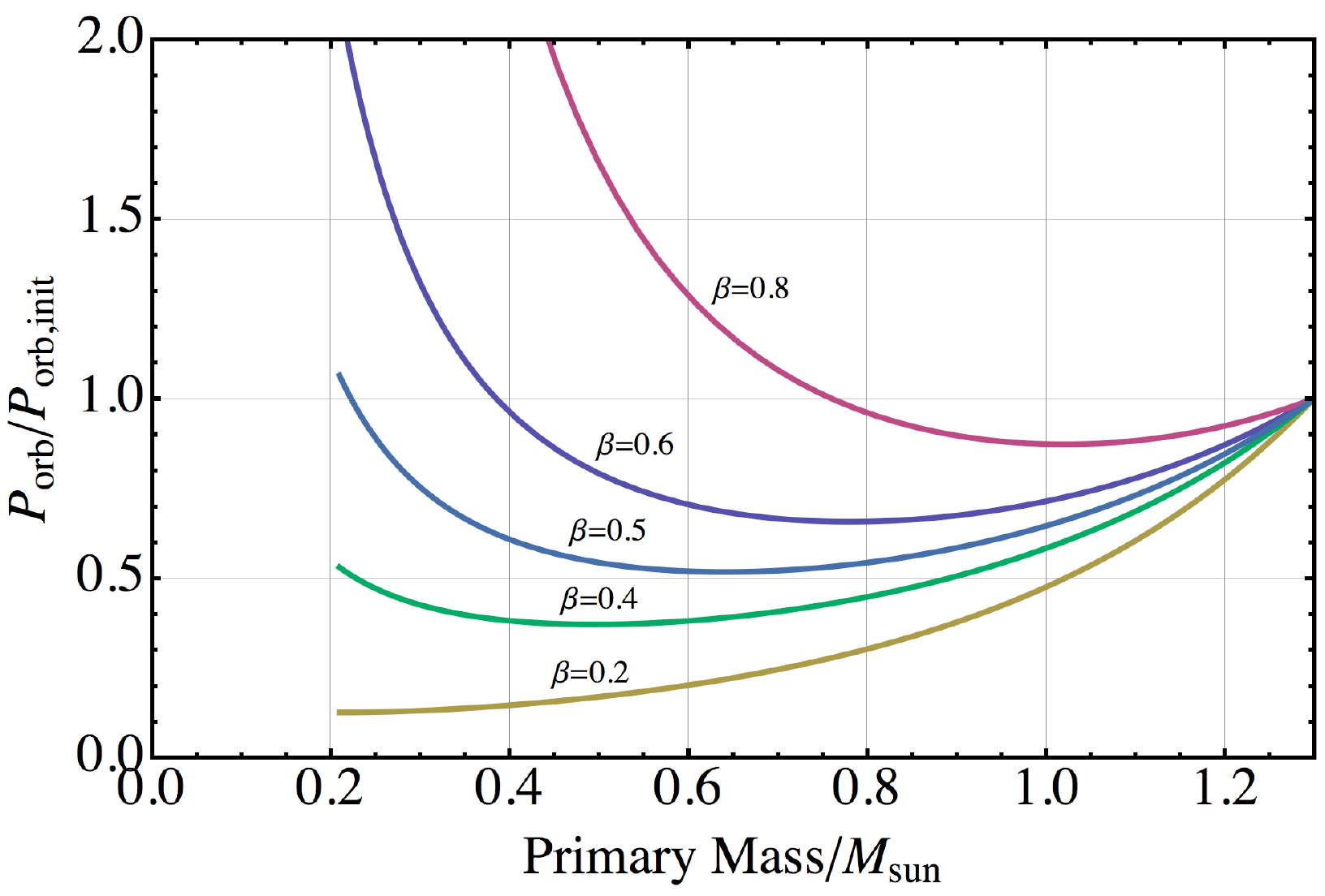}%
\else
    \includegraphics[width=0.9\hsize]{P_M.eps}%
\fi
\hfill}
\caption{Evolution of the orbital period of \KOI\ for different constant values of the mass retention fraction, $\beta$. In these evolutionary tracks, we assumed a single value of the angular momentum loss parameter, $\alpha = 1$. For illustrative purposes we chose the primordial masses to be $1.3\,M_\odot$ and $1.1\,M_\odot$. For $\beta \sim 0.5$, the orbital period remains constant to within a factor of two over the entire evolution. However, for $\beta \gtrsim 0.6$, the orbital period dramatically increases during the later portions of the binary evolution. Under the assumption that $\alpha$ and $\beta$ remain constant during the evolution, the mass transfer has to be relatively non-conservative, with $\beta$ probably no larger than $\sim 0.6$, in order to account for the short observed 2.7-day orbital period of \KOI. \label{f:orb_evolution}}
\end{figure}

Figure~\ref{f:orb_evolution} shows how the period of such a binary evolves as mass is transferred from the primary, for a range of plausible values of $\beta$. In all cases, the orbit shrinks as mass transfer gets under way. However, for $\beta \gtrsim 0.7$ the orbital period eventually dramatically increases above the original orbital period. Whereas, for smaller values of $\beta$ ($\lesssim 0.5$) the orbit either does not grow or continues to shrink with mass transfer. Regardless of the evolutionary path, the orbit cannot get close enough so that the accreting secondary overflows its Roche-lobe, otherwise the two stars would be in danger of merging.

\begin{figure}
\centerline{\hfill
\ifpdf
    \includegraphics[width=0.9\hsize]{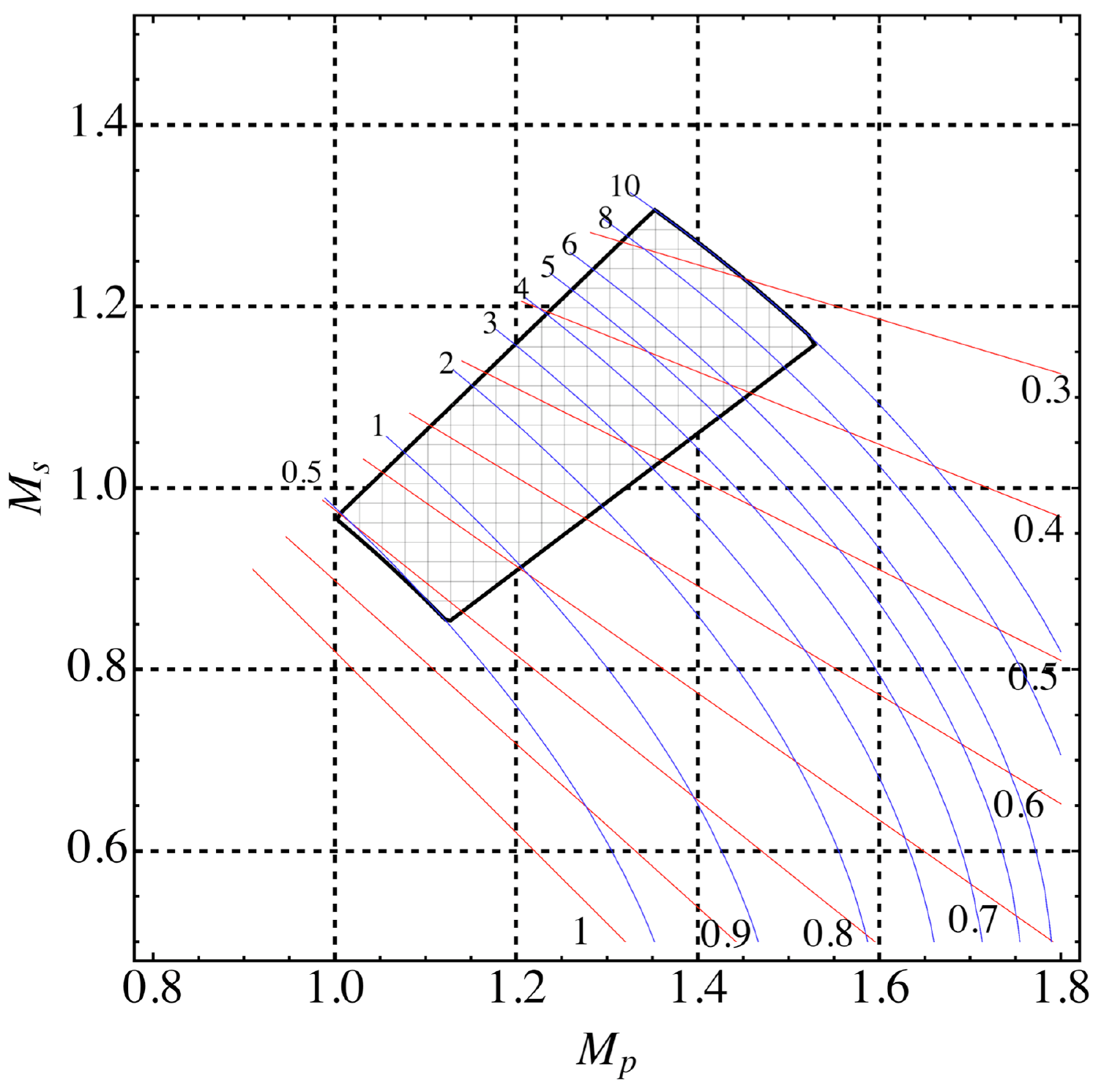}%
\else
    \includegraphics[width=0.9\hsize]{Mp_Ms_6606653.eps}%
\fi
\hfill}
\caption{Diagram showing the initial progenitor masses of a binary system that could evolve to become \KOI. Plotted are the initial primary (progenitor of the WD) mass, $M_p$, and the initial secondary (now the $\sim1.6\,M_\odot$ primary) mass, $M_s$. The red diagonal straight lines labeled $1 - 0.3$ (with a large font), are contours of the mass retention fraction, $\beta$ (i.e., the fraction of the mass leaving the primary that goes to the secondary and is not ejected from the system). The curved blue contours with values ranging from 0.5 to 10 (smaller font) are contours of constant initial orbital period (in days). The central shaded region provides the boundaries in the $M_p - M_s$ plane that would yield the observed masses set by constraints on the ratio of thermal and nuclear timescales and the requirement that the primordial primary underfills its initial Roche lobe by at least a factor of 1.3 (to allow it to evolve before transferring mass). The evolutionary tracks were calculated assuming that the specific angular momentum of the ejected mass in units of the angular momentum of the binary per unit of the binary reduced mass is $\alpha = 1$.\label{f:mp_ms}}
\end{figure}

The range of possible progenitor masses is explored in Figure~\ref{f:mp_ms}. The shaded region indicates the most plausible pairs of progenitor masses, $M_p$ and $M_s$. The constraints leading to this region are that (i) the ratio of nuclear and thermal timescales of the primordial pair lies between 0.5 and 0.9; (ii) the initial orbital period was shorter than $\sim$10 days; and (iii) the primary expanded by at least 30\% in radius by the time that Roche lobe overflow commenced. The red curves are contours of constant orbital period at the start of mass transfer, while the blue curves are contours of constant mass retention fraction, $\beta$. We again assumed a single value for the angular momentum loss parameter of $\alpha = 1$.

The rationale for constraint (i) above is that the secondary should have a nuclear evolution time that is not too much longer than that of the primary in order for it to have evolved off the main sequence only after the mass transfer was completed, but before the white dwarf had a chance to cool below $\sim$14,000\,K. The requirement (ii) for an initially short orbital period ($\lesssim 10$ days) is to avoid developing a white dwarf mass that is too large (i.e., remaining below $\sim 0.23 \, M_\odot$, our upper limit on $M_{\rm wd}$) to allow its progenitor to fit within its orbit at the time mass transfer commenced (see the discussion below). Finally, constraint (iii) allows room in the initial binary for the primary star to develop a He core before thermal timescale mass transfer reduces the primary's mass below $\sim$1\,$M_\odot$ and impedes its further evolution.

Finally, in regard to the evolutionary history of \KOI, we note that there is a theoretical relationship between the mass of the white dwarf and the orbital period in systems where the progenitor of the white dwarf was a low-mass star \citep[i.e., $\lesssim 2.2 \, M_\odot$; see, e.g.,][]{rappod95a,linrap11a} and where the mass transfer was stable. This is conveniently summarized in Figure~5 of \citet{linrap11a} and their equation (1). In their Figure~5 we can see that stars with initial mass $\lesssim 2.2 \,M_\odot$ (the red and green points in that plot) produce a steep functional relation between orbital period and the final core mass in the evolution of binary systems involving neutron star accretors. This same relation also somewhat holds for donor stars of initial mass up to $\sim$4 $M_\odot$ that commence mass transfer in so-called ``late case A" or ``case AB'' mass transfer, i.e., when the donor is near the end of its main-sequence phase. The fact that the accretors in the current problem are main sequence stars does not alter the core mass--radius relation on which this effect is based.

If we utilize the expression relating orbital period to core mass:
\begin{equation}
    P_{\rm orb} \simeq \frac{4.5 \times 10^6\,M_c^9}{\left(1+25M_c^{3.5}+29M_c^6\right)^{3/2}}~~{\rm days}
\end{equation}
\citep{linrap11a} we can infer that a final orbital period of 2.7 days should correspond to a white dwarf mass of $0.206 \, M_\odot$ with a spread in the theoretical values of $\sim \pm 0.015 \,M_\odot$. This is highly consistent with the mass of the white dwarf we find for \KOI. One caveat, however, is that the $P_{\rm orb}-M_{\rm wd}$ relation is based on the core mass--radius relation for giants \citep[see, e.g.][]{hanpod94a,rappod95a}.  In order for the latter to hold, the mass transfer must be effected in at least a semi-controlled way (see \S \ref{s:conclusion}).

\section{Conclusion}\label{s:conclusion}
In this paper we presented \KOI\ (KIC~6606653), a new {\em Kepler} eclipsing system consisting of a bloated white dwarf having a normal, slightly evolved companion star in a compact $2.698$-d orbit. Archival analysis of the optical and near-infrared photometry, as well as inspection of a preliminary classification spectrum (L. Nelson, private communication), reveal that the primary is likely an F6 spectral class $\sim 6350\,K$ star, while the {\em GALEX} UV flux allows us to infer a white dwarf temperature of $\sim 14,000\,K$, a value that is comparable with that found by our numerical modelling when using the primary's temperature as a prior.

We modelled the light curve of this binary with {\tt Icarus}, a new light curve synthesis code, which uses proper atmosphere models and a physical description of the binary parameters that allows us to reproduce various effects visible in the light curve such as irradiation, ellipsoidal light variations, and Doppler boosting. The derived binary parameters agree relatively well with those inferred from a harmonic analysis of the {\em Kepler} data using semi-analytic approximations of the above effects.

From our light curve fitting, we found masses of $1.59 \pm 0.07$ and $0.20 \pm 0.02\,M_\odot$ for the F-star primary and the white dwarf secondary, respectively, in a 2.698-day orbit. These values pose a modest challenge in terms of binary evolution. Indeed, the low mass of the white dwarf likely excludes a common envelope scenario, which would require a more evolved progenitor -- and therefore a more massive white dwarf -- in order to produce a compact system with the observed parameters. At the same time, the orbital separation of a system like \KOI\ is expected to increase significantly in a stable Roche-lobe overflow evolution unless the fraction of mass accreted by the current primary remains relatively low, $\beta \lesssim 0.5$. In any case, it appears that the system had to start its mass transfer evolution at an orbital period that is not longer than $\sim$10 days (as discussed above).

\KOI\ is the 4th such system found with {\em Kepler}.  The orbital periods are: 2.7 d (\KOI), 3.3 d  (KHWD3), 5.2 d (KOI~74), and 24 d (KOI~81) (current work; \citealt{carrap11a}; \citealt{vanrap10a}).  The corresponding white dwarf masses are $\sim$0.21, 0.26, 0.21, and 0.3 $M_\odot$, respectively, with $10-20\%$ uncertainties.  Unfortunately, the theoretical relationship between orbital period and white dwarf mass (discussed above) is too steep, i.e., $P_{\rm orb} \propto M^n_{\rm wd}$, where $n > 7$, to test effectively on such a relatively narrow range of $P_{\rm orb}$ and $M_{\rm wd}$. 

Preferably, one would like to find more such systems in {\em wider} orbital periods in the {\em Kepler} data where (i) the white dwarf masses are expected to deviate more substantially from $\sim$0.2 $M_\odot$, and (ii) the $P_{\rm orb}-M_{\rm wd}$ relation should work more robustly because of the higher likelihood of purely stable mass transfer.  In this regard, we note that \citet{maxand11a} have recently discovered a possibly related system with an orbital period of only 0.688 days, and a highly bloated white dwarf of $0.29 \pm 0.005 \, M_\odot$, as well as several other similar very short period systems (P. Maxted, private communication).  The $P_{\rm orb}-M_{\rm wd}$ relation is difficult to quantify theoretically at these very short periods (of $\lesssim$ 1 day) because it is not yet clear whether these systems result from stable mass transfer, a common envelope scenario, or some less catastrophic form of unstable mass transfer.  The $P_{\rm orb}-M_{\rm wd}$ relation works only when mass is removed from the progenitor of the white dwarf in a reasonably controlled way such that the core-mass--radius relation, on which it is based, remains valid.

\acknowledgments

The authors are grateful to the Canadian Institute for Theoretical Astrophysics for the use of the Sunnyvale computer cluster as well as Lorne Nelson for sharing his early spectroscopic observation of \KOI. We acknowledge use of CDS's VizieR and Simbad and NASA's ADS and MAST services. RPB thanks Mubdi Rahman for insightfully discussions about Python while developing the {\tt Icarus} code as well as France Allard for support with the Phoenix model atmosphere simulator.

{\it Facilities:} \facility{{\em Kepler}}, \facility{{\em GALEX}}.

\bibliography{kic_6606653}

\begin{thebibliography}{57}
\expandafter\ifx\csname natexlab\endcsname\relax\def\natexlab#1{#1}\fi

\bibitem[{{Allard} {et~al.}(2007){Allard}, {Allard}, {Homeier}, {Kielkopf},
  {McCaughrean}, \& {Spiegelman}}]{allall07a}
{Allard}, F., {Allard}, N.~F., {Homeier}, D., {Kielkopf}, J., {McCaughrean},
  M.~J., \& {Spiegelman}, F. 2007, \aap, 474, L21

\bibitem[{{Allard} {et~al.}(2003){Allard}, {Guillot}, {Ludwig}, {Hauschildt},
  {Schweitzer}, {Alexander}, \& {Ferguson}}]{allgui03a}
{Allard}, F., {Guillot}, T., {Ludwig}, H.-G., {Hauschildt}, P.~H.,
  {Schweitzer}, A., {Alexander}, D.~R., \& {Ferguson}, J.~W. 2003, in IAU
  Symposium, Vol. 211, Brown Dwarfs, ed. {E.~Mart{\'{\i}}n}, 325--+

\bibitem[{{Allard} {et~al.}(2010){Allard}, {Homeier}, \& {Freytag}}]{allhom10a}
{Allard}, F., {Homeier}, D., \& {Freytag}, B. 2010, ArXiv e-prints

\bibitem[{{Avni} \& {Bahcall}(1975)}]{avnbah75a}
{Avni}, Y., \& {Bahcall}, J.~N. 1975, \apj, 197, 675

\bibitem[{{Beech}(1985)}]{bee85a}
{Beech}, M. 1985, \apss, 117, 69

\bibitem[{{Bianchi} {et~al.}(2007)}]{bian+07}
{Bianchi}, L., {et~al.} 2007, \apjs, 173, 659

\bibitem[{{Borucki} {et~al.}(2011){Borucki}, {Koch}, {Basri}, {Batalha},
  {Brown}, {Bryson}, {Caldwell}, {Christensen-Dalsgaard}, {Cochran}, {DeVore},
  {Dunham}, {Gautier}, {Geary}, {Gilliland}, {Gould}, {Howell}, {Jenkins},
  {Latham}, {Lissauer}, {Marcy}, {Rowe}, {Sasselov}, {Boss}, {Charbonneau},
  {Ciardi}, {Doyle}, {Dupree}, {Ford}, {Fortney}, {Holman}, {Seager},
  {Steffen}, {Tarter}, {Welsh}, {Allen}, {Buchhave}, {Christiansen}, {Clarke},
  {Das}, {D{\'e}sert}, {Endl}, {Fabrycky}, {Fressin}, {Haas}, {Horch},
  {Howard}, {Isaacson}, {Kjeldsen}, {Kolodziejczak}, {Kulesa}, {Li}, {Lucas},
  {Machalek}, {McCarthy}, {MacQueen}, {Meibom}, {Miquel}, {Prsa}, {Quinn},
  {Quintana}, {Ragozzine}, {Sherry}, {Shporer}, {Tenenbaum}, {Torres},
  {Twicken}, {Van Cleve}, {Walkowicz}, {Witteborn}, \& {Still}}]{borkoc11a}
{Borucki}, W.~J., {et~al.} 2011, \apj, 736, 19

\bibitem[{{Brown} {et~al.}(2011){Brown}, {Latham}, {Everett}, \&
  {Esquerdo}}]{brow+11}
{Brown}, T.~M., {Latham}, D.~W., {Everett}, M.~E., \& {Esquerdo}, G.~A. 2011,
  arXiv:1102.0342

\bibitem[{{Carter} {et~al.}(2011){Carter}, {Rappaport}, \&
  {Fabrycky}}]{carrap11a}
{Carter}, J.~A., {Rappaport}, S., \& {Fabrycky}, D. 2011, \apj, 728, 139

\bibitem[{{Che} {et~al.}(2011){Che}, {Monnier}, {Zhao}, {Pedretti}, {Thureau},
  {M{\'e}rand}, {ten Brummelaar}, {McAlister}, {Ridgway}, {Turner}, {Sturmann},
  \& {Sturmann}}]{chemon11a}
{Che}, X., {et~al.} 2011, \apj, 732, 68

\bibitem[{{Covey} {et~al.}(2007)}]{cove+07}
{Covey}, K.~R., {et~al.} 2007, \aj, 134, 2398

\bibitem[{{Cox}(2000)}]{cox00}
{Cox}, A.~N. 2000, {Allen's astrophysical quantities}, ed. {Cox, A.~N.}

\bibitem[{{de Kool}(1990)}]{dek90a}
{de Kool}, M. 1990, \apj, 358, 189

\bibitem[{{Dewi} \& {Tauris}(2000)}]{dewtau00a}
{Dewi}, J.~D.~M., \& {Tauris}, T.~M. 2000, \aap, 360, 1043

\bibitem[{{Di Stefano}(2011)}]{dist11a}
{Di Stefano}, R. 2011, \aj, 141, 142

\bibitem[{{Driebe} {et~al.}(1999){Driebe}, {Bl{\"o}cker}, {Sch{\"o}nberner}, \&
  {Herwig}}]{driblo99a}
{Driebe}, T., {Bl{\"o}cker}, T., {Sch{\"o}nberner}, D., \& {Herwig}, F. 1999,
  \aap, 350, 89

\bibitem[{{Farmer} \& {Agol}(2003)}]{farm03a}
{Farmer}, A.~J., \& {Agol}, E. 2003, \apj, 592, 1151

\bibitem[{{Fukugita} {et~al.}(1996){Fukugita}, {Ichikawa}, {Gunn}, {Doi},
  {Shimasaku}, \& {Schneider}}]{fuku+96}
{Fukugita}, M., {Ichikawa}, T., {Gunn}, J.~E., {Doi}, M., {Shimasaku}, K., \&
  {Schneider}, D.~P. 1996, \aj, 111, 1748

\bibitem[{Gelman \& Rubin(1992)}]{gelm92a}
Gelman, A., \& Rubin, D.~B. 1992, Statistical Science, 7, 457

\bibitem[{{Gies} {et~al.}(2008){Gies}, {Dieterich}, {Richardson}, {Riedel},
  {Team}, {McAlister}, {Bagnuolo}, {Grundstrom}, {{\v S}tefl}, {Rivinius}, \&
  {Baade}}]{giedie08a}
{Gies}, D.~R., {et~al.} 2008, \apjl, 682, L117

\bibitem[{{Gray} {et~al.}(2001){Gray}, {Graham}, \& {Hoyt}}]{gragra01a}
{Gray}, R.~O., {Graham}, P.~W., \& {Hoyt}, S.~R. 2001, \aj, 121, 2159

\bibitem[{{Gregory}(2005)}]{gre05b}
{Gregory}, P.~C. 2005, {Bayesian Logical Data Analysis for the Physical
  Sciences: A Comparative Approach with `Mathematica' Support}, ed. {Gregory,
  P.~C.} (Cambridge University Press)

\bibitem[{{Han} {et~al.}(1994){Han}, {Podsiadlowski}, \&
  {Eggleton}}]{hanpod94a}
{Han}, Z., {Podsiadlowski}, P., \& {Eggleton}, P.~P. 1994, \mnras, 270, 121

\bibitem[{{Hendry} \& {Mochnacki}(1992)}]{henmoc92a}
{Hendry}, P.~D., \& {Mochnacki}, S.~W. 1992, \apj, 388, 603

\bibitem[{{Kippenhahn} \& {Meyer-Hofmeister}(1977)}]{kipmey77a}
{Kippenhahn}, R., \& {Meyer-Hofmeister}, E. 1977, \aap, 54, 539

\bibitem[{{Kopal}(1959)}]{kop59a}
{Kopal}, Z. 1959, {Close binary systems}, ed. {Kopal, Z.}

\bibitem[{{Lin} {et~al.}(2011){Lin}, {Rappaport}, {Podsiadlowski}, {Nelson},
  {Paxton}, \& {Todorov}}]{linrap11a}
{Lin}, J., {Rappaport}, S., {Podsiadlowski}, P., {Nelson}, L., {Paxton}, B., \&
  {Todorov}, P. 2011, \apj, 732, 70

\bibitem[{{Loeb} \& {Gaudi}(2003)}]{loegau03a}
{Loeb}, A., \& {Gaudi}, B.~S. 2003, \apjl, 588, L117

\bibitem[{{Lucy}(1967)}]{luc67a}
{Lucy}, L.~B. 1967, \zap, 65, 89

\bibitem[{{Maxted} {et~al.}(2011){Maxted}, {Anderson}, {Burleigh},
  {Collier-Cameron}, {Heber}, {Gaensicke}, {Geier}, {Kupfer}, {Marsh},
  {Nelemans}, {O'Toole}, {Ostensen}, {Smalley}, \& {West}}]{maxand11a}
{Maxted}, P.~F.~L., {et~al.} 2011, ArXiv e-prints 1107.4986v2

\bibitem[{{McLaughlin}(1924)}]{mcl24a}
{McLaughlin}, D.~B. 1924, \apj, 60, 22

\bibitem[{{Morris}(1985)}]{mor85a}
{Morris}, S.~L. 1985, \apj, 295, 143

\bibitem[{{Neckel}(2005)}]{nec05a}
{Neckel}, H. 2005, \solphys, 229, 13

\bibitem[{{Nelson} {et~al.}(2004){Nelson}, {Dubeau}, \&
  {MacCannell}}]{neldub04a}
{Nelson}, L.~A., {Dubeau}, E., \& {MacCannell}, K.~A. 2004, \apj, 616, 1124

\bibitem[{{Orosz} \& {Hauschildt}(2000)}]{orohau00a}
{Orosz}, J.~A., \& {Hauschildt}, P.~H. 2000, \aap, 364, 265

\bibitem[{{Paxton} {et~al.}(2011){Paxton}, {Bildsten}, {Dotter}, {Herwig},
  {Lesaffre}, \& {Timmes}}]{paxbil11a}
{Paxton}, B., {Bildsten}, L., {Dotter}, A., {Herwig}, F., {Lesaffre}, P., \&
  {Timmes}, F. 2011, \apjs, 192, 3

\bibitem[{{Podsiadlowski} {et~al.}(2003){Podsiadlowski}, {Rappaport}, \&
  {Han}}]{podrap03a}
{Podsiadlowski}, P., {Rappaport}, S., \& {Han}, Z. 2003, \mnras, 341, 385

\bibitem[{{Podsiadlowski} {et~al.}(2002){Podsiadlowski}, {Rappaport}, \&
  {Pfahl}}]{podrap02a}
{Podsiadlowski}, P., {Rappaport}, S., \& {Pfahl}, E.~D. 2002, \apj, 565, 1107

\bibitem[{{Pr{\v s}a} \& {Zwitter}(2005)}]{prszwi05a}
{Pr{\v s}a}, A., \& {Zwitter}, T. 2005, \apj, 628, 426

\bibitem[{{Pr{\v s}a} {et~al.}(2011){Pr{\v s}a}, {Batalha}, {Slawson}, {Doyle},
  {Welsh}, {Orosz}, {Seager}, {Rucker}, {Mjaseth}, {Engle}, {Conroy},
  {Jenkins}, {Caldwell}, {Koch}, \& {Borucki}}]{prsbat11a}
{Pr{\v s}a}, A., {et~al.} 2011, \aj, 141, 83

\bibitem[{{Rappaport} {et~al.}(2009){Rappaport}, {Podsiadlowski}, \&
  {Horev}}]{rappod09a}
{Rappaport}, S., {Podsiadlowski}, P., \& {Horev}, I. 2009, \apj, 698, 666

\bibitem[{{Rappaport} {et~al.}(1995){Rappaport}, {Podsiadlowski}, {Joss}, {Di
  Stefano}, \& {Han}}]{rappod95a}
{Rappaport}, S., {Podsiadlowski}, P., {Joss}, P.~C., {Di Stefano}, R., \&
  {Han}, Z. 1995, \mnras, 273, 731

\bibitem[{{Rey} {et~al.}(2007)}]{rey+07}
{Rey}, S.-C., {et~al.} 2007, \apjs, 173, 643

\bibitem[{{Rossiter}(1924)}]{ros24a}
{Rossiter}, R.~A. 1924, \apj, 60, 15

\bibitem[{{Rowe} {et~al.}(2010){Rowe}, {Borucki}, {Koch}, {Howell}, {Basri},
  {Batalha}, {Brown}, {Caldwell}, {Cochran}, {Dunham}, {Dupree}, {Fortney},
  {Gautier}, {Gilliland}, {Jenkins}, {Latham}, {Lissauer}, {Marcy}, {Monet},
  {Sasselov}, \& {Welsh}}]{rowbor10a}
{Rowe}, J.~F., {et~al.} 2010, \apjl, 713, L150

\bibitem[{{Schlegel} {et~al.}(1998){Schlegel}, {Finkbeiner}, \&
  {Davis}}]{schlfd98}
{Schlegel}, D.~J., {Finkbeiner}, D.~P., \& {Davis}, M. 1998, \apj, 500, 525

\bibitem[{{Shporer} {et~al.}(2011){Shporer}, {Brown}, {Mazeh}, \&
  {Zucker}}]{shpbro11a}
{Shporer}, A., {Brown}, T., {Mazeh}, T., \& {Zucker}, S. 2011, ArXiv e-prints

\bibitem[{{Sing}(2010)}]{sin10a}
{Sing}, D.~K. 2010, \aap, 510, A21+

\bibitem[{{Skrutskie} {et~al.}(2006){Skrutskie}, {Cutri}, {Stiening},
  {Weinberg}, {Schneider}, {Carpenter}, {Beichman}, {Capps}, {Chester},
  {Elias}, {Huchra}, {Liebert}, {Lonsdale}, {Monet}, {Price}, {Seitzer},
  {Jarrett}, {Kirkpatrick}, {Gizis}, {Howard}, {Evans}, {Fowler}, {Fullmer},
  {Hurt}, {Light}, {Kopan}, {Marsh}, {McCallon}, {Tam}, {Van Dyk}, \&
  {Wheelock}}]{skru+06}
{Skrutskie}, M.~F., {et~al.} 2006, \aj, 131, 1163

\bibitem[{{Tauris} \& {Dewi}(2001)}]{taudew01a}
{Tauris}, T.~M., \& {Dewi}, J.~D.~M. 2001, \aap, 369, 170

\bibitem[{{van Kerkwijk} {et~al.}(2010){van Kerkwijk}, {Rappaport}, {Breton},
  {Justham}, {Podsiadlowski}, \& {Han}}]{vanrap10a}
{van Kerkwijk}, M.~H., {Rappaport}, S.~A., {Breton}, R.~P., {Justham}, S.,
  {Podsiadlowski}, P., \& {Han}, Z. 2010, \apj, 715, 51

\bibitem[{{van Rensbergen} {et~al.}(2010){van Rensbergen}, {De Greve},
  {Mennekens}, {Jansen}, \& {De Loore}}]{vandeg10a}
{van Rensbergen}, W., {De Greve}, J.~P., {Mennekens}, N., {Jansen}, K., \& {De
  Loore}, C. 2010, \aap, 510, A13+

\bibitem[{{van Rensbergen} {et~al.}(2011){van Rensbergen}, {de Greve},
  {Mennekens}, {Jansen}, \& {de Loore}}]{vandeg11a}
{van Rensbergen}, W., {de Greve}, J.~P., {Mennekens}, N., {Jansen}, K., \& {de
  Loore}, C. 2011, \aap, 528, A16+

\bibitem[{{Vennes} {et~al.}(2011){Vennes}, {Kawka}, \& {N{\'e}meth}}]{venn+11}
{Vennes}, S., {Kawka}, A., \& {N{\'e}meth}, P. 2011, \mnras, 410, 2095

\bibitem[{{von Zeipel}(1924)}]{von24a}
{von Zeipel}, H. 1924, \mnras, 84, 665

\bibitem[{Walsh(2004)}]{wals2004}
Walsh, B. 2004, Markov Chain Monte Carlo and Gibbs Sampling

\bibitem[{{Wilson} \& {Devinney}(1971)}]{wildev71a}
{Wilson}, R.~E., \& {Devinney}, E.~J. 1971, \apj, 166, 605

\end{thebibliography}

\newpage

\appendix
\section{{\tt Icarus}: Binary Light Curve Synthesis Code}\label{s:icarus}
\subsection{Stellar Surface Grid}
The underlying stellar surface grid is constructed using the triangular tessellation of a unit sphere that was inspired by the {\tt GDDSYN} synthesis code \citep{henmoc92a}. The base vertices are initially generated from the primitives of an icosahedron. Each triangular surface is then sub-divided into four triangles by creating a new vertex at the midpoint of each edge. The new vertices are finally projected back onto the unit sphere by re-normalizing them to unity. This sub-division process can be repeated iteratively in order to obtain the desired number of surface elements. At a sub-division level $n$ (where $n=0$ is the base icosahedron, and $n=1$ is one sub-division further), the number of surface elements is $n_{\rm faces}=20\, (4^n)$, while the number of independent vertices is $n_{\rm vertices}=2+10\,(4^n)$. The coordinates of each face is calculated as the centroid of the triangle.

There are several advantages to using the algorithm above. First, a triangular tessellation does not suffer from the uneven surface sampling problem common to simple meshes derived from an equally spaced grid in spherical coordinates. To circumvent this problem, other synthesis codes, like {\tt PHOEBE}, use a variable sampling strategy with fewer points near the poles. The subdivision algorithm of a triangular tessellation is relatively easy to implement and, using an icosahedron as the primitive, yields triangular faces that are organized on a regular tile of hexagons and pentagons, and that have fairly equal areas. Triangular faces are by definition convex and hence are simpler to work with when it comes to dealing with transits and eclipses. For reasons that will become obvious later we also keep track of the associativities: each surface is associated to three vertices, and each vertex is associated to either 5 or 6 faces.

\subsection{Equipotential Surface}\label{s:equipotential}
As in the ELC code, our stellar surface is defined by the gravitational equipotential equation from \citet{avnbah75a}, which takes into account the effects of stellar rotation:
\begin{equation}
    \psi = \frac{G M_1}{a} \left[ \frac{1}{r_1} + \frac{q}{r_2} - q x + \frac{q+1}{2} \Omega^2 \left(r_1^2-z^2\right) \right] \,,
\end{equation}
where $r_1$ and $r_2$ is the distance of a point measured from the barycenter of the star that is being modelled (labeled 1) and from the barycenter of its companion (labeled 2) in units of orbital separation, $q = M_2/M_1$ is the mass ratio, and $\Omega = \omega_{\rm star}/\omega_{\rm orbit}$ is the co-rotation factor expressed as the ratio of the stellar to orbital frequency. We work in a coordinate system having its origin at the barycenter of star 1, in which the $x$-axis points towards star 2's barycenter, the $z$-axis is along the orbital angular momentum, and the $y$-axis is along the orbital plane orthogonal to the $x$ and $z$ axes. Our code is currently limited to circular orbits but could easily be extended to elliptical orbits as in {\tt PHOEBE}, for example.

Initially, the position of the L1 point, $x_{\rm L1}$, along the x-axis is found by identifying the saddle-point of the above equation between stars 1 and 2. Then, as explained in \citet{orohau00a}, the potential corresponding to the ``nose'' of the star (the inner point towards star 2) is defined by specifying a filling factor $f = x_{\rm nose}/x_{\rm L1}$, which is expressed as a fraction of the L1 distance. From there, the radius of star 1 can be evaluated in the direction of every vertex and face of the tessellated surface.

The effective surface gravity is also calculated for each face using the equation $g = \lVert \vec\nabla\psi \rVert$ as well as the components of the normal to the local surface.

\subsection{Surface Temperature and Gravity Darkening}
We assign a fiducial temperature to each face that takes into account the gravity darkening according to the equation:
\begin{equation}
    \frac{T(x,y,z)}{T_{\rm pole}} = \left[ \frac{g(x,y,z)}{g_{\rm pole}} \right]^\beta \,,
\end{equation}
where $\beta$ typically varies between 0.08 and 0.25 depending on whether the star has a convective or a radiative envelope, respectively (\citealt{luc67a,von24a}; for empirical constraints, see \citealt{chemon11a}).

\subsection{Irradiation from a Companion}\label{s:irradiation}
We treat the effect of irradiation from a companion as a source of energy that increases the temperature of the star while preserving thermal equilibrium (i.e., no thermal inversion that gives rises to emission lines). In this case, irradiation changes the temperature distribution of the star as following:
\begin{equation}
    T^\prime(x,y,z) = \left[ T(x,y,z)^4 + T_{\rm irr}^4 \frac{\cos \chi}{r_2^2} \right]^{1/4} \,,
\end{equation}
\citep{bee85a} where $\cos \chi$ is the angle between the surface normal and the direction to the companion (i.e., star 2), and $T_{\rm irr}$ is the irradiation temperature which describes the temperature increase. Note that working with $T_{\rm irr}$ is more convenient than using an irradiation luminosity and a bolometric albedo since empirically one effectively measures the back and front temperatures of a star using multi-band light curves. One can convert the irradiation temperature into an irradiation luminosity after the fact using $L_{\rm irr} = 4\pi\sigma a^2 T_{\rm irr}^4$, where $\sigma$ is the Stefan-Boltzmann constant and $a$ the orbital separation, and work out the stellar albedo if the companion's luminosity is known such that $\alpha = L_{\rm irr} / L_2$. If one prefers, it is also possible to fix an albedo and perform the calculation the other way around in order to provide an irradiation temperature to the code.

Our treatment of irradiation is different than that of {\tt PHOEBE} and {\tt GDDSYN}, which calculate the reflection effect of each surface element due to the other star in an iterative way. While their technique is certainly more accurate, it comes at the cost of a substantially larger computational burden and only becomes more relevant in the case of contact binaries, where one needs to account for shadowing effects due to the bridge of matter in order to be self-consistent. Moreover, the fact that heat might be partly redistributed over the surface in a way that is non-trivial to calculate from first principles helps justify our use of this approximation.

\subsection{Surface Area}
We pre-calculate the area of each surface triangle on the unit sphere. Once the equipotential equation has been solved we rescale each pre-calculated value by its proper radius in order to obtain the effective area of the surface elements. By this means, the integrated flux simply corresponds to a discrete summation over the visible surface.

\subsection{Atmosphere Models}\label{s:atmosphere}
Once the stellar grid has been constructed, and the effective temperature and surface gravity determined for all the faces, we can evaluate the flux perceived by an observer located at a given orbital inclination and orbital phase. The backend of the code that returns the flux for each face can be swapped between a blackbody or an atmosphere grid.

For the purpose of this paper, we have worked with the BTSettl atmosphere models of \citet{allgui03a,allall07a,allhom10a} which are available online on the Phoenix web simulator\footnote{Available at \url{http://phoenix.ens-lyon.fr/simulator/}}. We used these atmosphere models to cover a grid over the range 1000-15000\,K in temperature and 3.0-4.5 in $\log g$. Since only integrated spectra were available, we used the empirical limb darkening relationship of \citet{nec05a}.

Prior to performing the modelling of the \KOI\ data, we integrated the spectral grid over the {\em Kepler} passband. Such integration can be performed for any photometric filter that one wishes to use to model the light curves. However, our code can also handle working with a full set of spectral data. In which case it is possible to model full orbital-resolved spectroscopic data. When this is the case, the spectrum of each face is Doppler shifted at the appropriate velocity and hence the rotational broadening, orbital Doppler shift, and displacement of the light-center away from the barycenter due to irradiation naturally arise from our model spectra. In principle, the Rossiter-McLaughlin effect \citep{ros24a,mcl24a} (as well as its Doppler boosting analogue \citep{vanrap10a,shpbro11a}) would also appear from spectra modelled in combination with the transit calculation described below.

\subsection{Doppler Boosting}\label{s:doppler}
Doppler boosting is another feature implemented in our code. For most cases, the amplitude of the effect is negligible, and hence it is turned off to speed up the light curve calculation. For the \KOI\ data, however, Doppler boosting is clearly detected and hence it was included in the following way.

As described in \citet{vanrap10a}, Doppler boosting can be written as a modulation of the observed flux:
\begin{equation}\label{e:doppler}
    f = 1 - f_{\rm DB} \sin\phi\,v/c \,,
\end{equation}
where $\phi$ is the orbital phase measured at the inferior conjunction of the star, $v$ is the orbital velocity, $c$ is the speed of light, and $f_{\rm DB}$ is a factor that depends on the spectrum of the source and the observing passband. Based on our analytical estimate of the primary's temperature and surface gravity (see \S\ref{s:galex}), we evaluated the coefficient $f_{\rm DB}$ at temperatures of 6200 and 6500\,K for $\log g$ of 3.5 and 4.0. The coefficient is a rather smooth function of the temperature of surface gravity, and hence we simply performed a bilinear interpolation to fetch the coefficient to apply to each face of the stellar grid before the total flux was integrated. In the future, we are planning to include the exact calculation at each temperature and $\log g$ value.

\subsection{Eclipse and Transit Calculation}
Eclipse and transit calculation are also included in our synthesis code. Each time that the stellar surface grid is recomputed with new input parameters, we perform several tests to evaluate whether partial or total occultations are possible as well as the orbital phase range within which they can occur. This allows for our code to revert to algorithms optimized for each situation.

At an orbital phase $\phi$ and an inclination $i$, the projected distance between the barycenter of the stars is given by $\left(\delta_a\right)^2 = a^2 \left( \cos^2 i \cos^2 \phi + \sin^2 i \right)$, where $\phi=0$ at the inferior conjunction of the primary.

First, we test whether any kind of occultation can occur. We consider the maximum dimension of each star in order to obtain a conservative estimate. Hence, for any orbital range where $\delta_a < r_{1,{\rm max}} + r_{2,{\rm max}}$, one star might be  occultated by the other. If this is the case, a second test aims to check if a full occultation occurs, and, if so, what orbital range does it cover. If $r_{1,{\rm max}} < r_{2,{\rm min}}$, star 1 will definitely get fully eclipsed in the orbital range that meets the requirement $\delta_a + r_{1,{\rm max}} < r_{2,{\rm min}}$. The same test is also performed with the indices reversed.

With these criteria identified, the flux calculation can be classified into three categories: (1) no occultation, (2) partial occultation, and (3) full occultation. Case 3 is the simplest as the star does not contribute to the total observed flux and hence no calculation is required. Case 1 is also relatively simple as every face that is not beyond the stellar limb will contribute; a standard flux integration over the visible surface of each star is performed. Case 2 is more complex and we explain our algorithm below.

\subsubsection{Partial Visibility}
Evaluating the partial visibility of a stellar surface  occultated by an other is a computationally intensive task that can be tackled in several ways. Our algorithm is meant to provide a balance between accuracy, execution speed, and ease of implementation.

In short summary, our algorithm aims to calculate the fraction of a surface element that is visible to an observer. The first step consists in the determination of the outline of the star located in front (here we choose star 2 for the explanation). We make the approximation that the orbital range within which the occultation of star 1 occurs is small enough that the outline at the onset of the occultation is the same as at conjunction. This simplifies the calculation considerably since, when the orbital inclination is sufficiently close to edge-on\footnote{A more accurate outline could be calculated but our approximation is good enough in the case of \KOI}, the limb of star 2 corresponds to its outline in the $y-z$ plane at $x=0$, which is easy to compute. In this fashion, we obtain a lookup table of $r_2(\theta)$ values, where $\theta$ and $r_2$ are the polar coordinates of the limb of star 2 in the sky plane calculated using its barycenter as the origin.

In the second phase, we compute the projected position of star 1's vertices, $\theta_1$ and $r_1$, with respect to star 2's barycenter. If a vertex of star 1 lies within the limb of star 2 -- that is $r_1 < r_2(\theta_1)$ -- then it is hidden. To calculate the partial visibility of a surface element, we make the approximation the visible fraction of the surface is equal to the fraction of its vertices that are not hidden. Since each individual vertex is associated with either five or six faces, we can use the vertex-to-face associativity to assign the partial visibility weighting factor to all the faces without having to perform redundant calculations.

\begin{figure*}
\centerline{\hfill
\ifpdf
    \includegraphics[width=0.5\hsize]{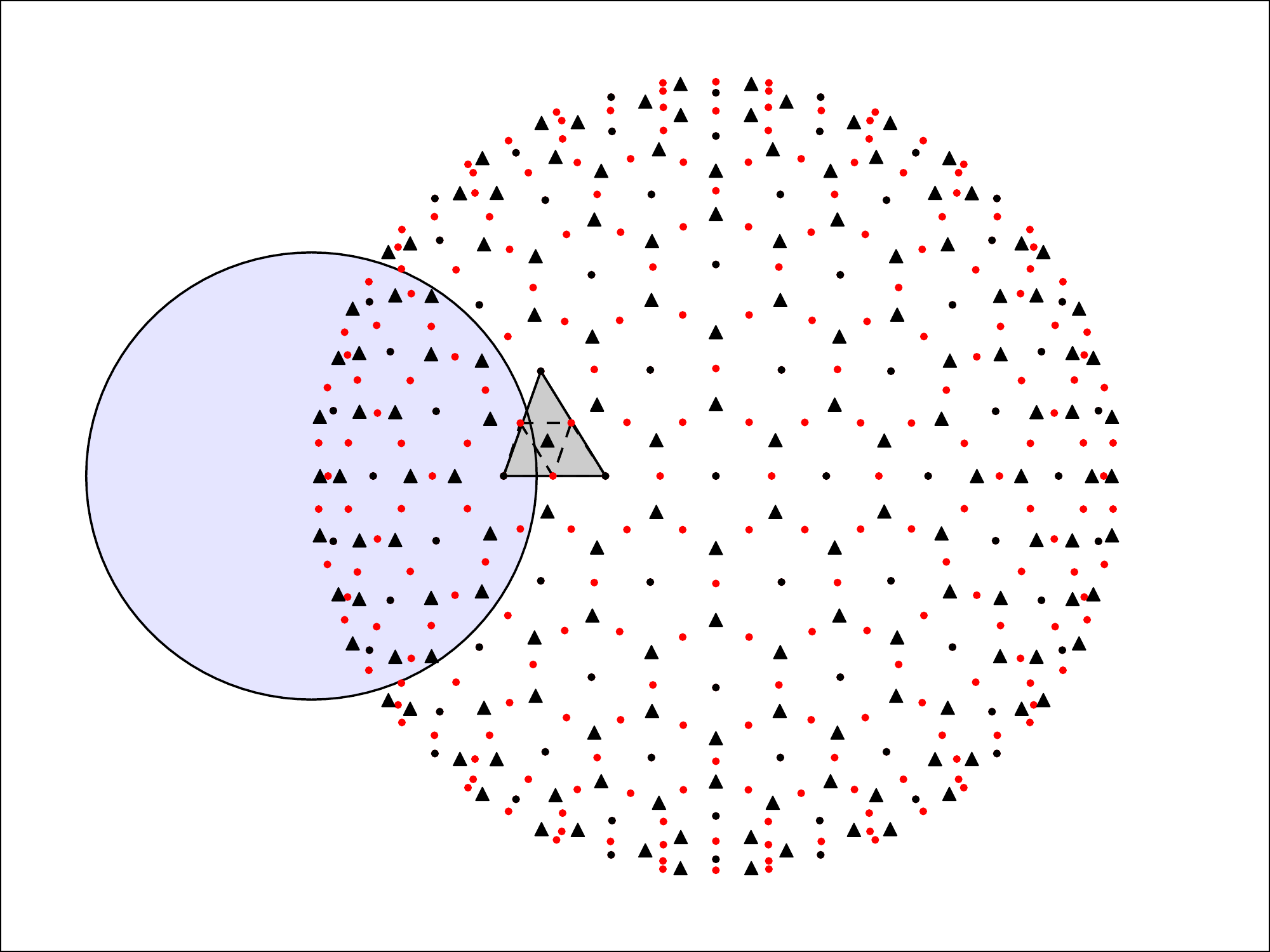}%
\else
    \includegraphics[width=0.5\hsize]{schematic_surface.pdf}%
\fi
\hfill}
\caption{Diagram illustrating an occultation of the primary (larger star) by the secondary (smaller, blue star). The primary's tessellated surface is displayed with the following details: the triangle symbols mark the center of the triangular surfaces, the black dots are the vertices of the triangular surfaces and the red dots are the vertices of nested triangular surface sub-divisions. To evaluate the visibility of the surface elements, the outline of the secondary is found and a lookup table $r_2({\theta})$ created, where $\theta$ and $r_2$ are the polar coordinates of the limb in the sky plane, using the secondary's barycenter as origin. The projected distance of each vertex from the secondary's barycenter is then calculated. Surface elements having all their vertices lying within the light blue region are not visible. For those near the boundary, a weighting is applying. For the surface element highlighted in grey, one of its primary vertices is occulted and so is one of its sub-vertices. Hence, $2/6$ vertices are hidden which implies that this surface element only contributes $33\%$ its normal flux.\label{f:schematic}}
\end{figure*}

Our partial visibility algorithm relies on a simple projected distance determination for all the vertices of the visible face. When such a vertex is occulted, an extra weighting factor is added to five or six array elements corresponding to the associated surface elements. The computational cost of this technique remains small compared to a more accurate calculation such as that performed by {\tt GDDSYN}. Moreover it can easily be parallelized.

The main caveat of our algorithm lies in the limited spatial resolution of the grid faces; the partial visibility goes in $1/3$ increments. Hence, the light curve can display important jitter if the chosen grid resolution is too coarse. A notable symptomatic case is that of stars having very unequal relative sizes and for which the projected area of the smaller star becomes comparable to the area of a surface elements of the larger star. The obvious solution relies on using a higher resolution for the larger star, though this comes at the expense of an increased computational burden.

To work around these problems, we have designed a method that takes advantage of the sub-division tessellation algorithm. Since the sub-division to a finer level adds vertices at the mid-points of the surface sides, it is easy to keep track of the associations of these vertices with the parent surface elements at the lower resolution. Using such a nested sub-division yields partial visibilities that can be computed to a much better precision than the original $1/3$ step-size, increasing to $1/6$ for one extra sub-division and $1/15$ for two. This algorithm comes with a relatively small computing footprint: since the surface properties (temperature and gravity) vary smoothly, it is sufficient to obtain an accurate partial visibility weighting for a low-resolution grid using the sub-division. In other words, there is little if no gain in calculating the emergent intensity for all the surface elements at the higher resolution.

\end{document}